\documentclass[pre,epsfig,address,twocolumn]{revtex4}%
\usepackage[english]{babel}
\usepackage[latin1]{inputenc}
\usepackage[dvips]{graphicx}
\usepackage{amsmath}
\usepackage{amssymb}
\usepackage{epsfig}
\usepackage{array}
\usepackage{amsfonts}
\usepackage{graphicx}%
\begin{document}
\title{Asymmetric simple exclusion process with periodic boundary driving }

\author{Vladislav Popkov$^{1,2}$, Mario Salerno$^{1}$ and Gunter M. Sch\"utz$^{2,3}$}
\affiliation{ $^1$ Dipartimento di Fisica "E.R. Caianiello", and
Consorzio Nazionale Interuniversitario per le Scienze Fisiche
della Materia (CNISM), Universit\`a di Salerno, Baronissi,
Italy\\
$^2$ Interdisziplin\"ares Zentrum fur Komplexe Systeme,
R\"omerstrasse
164, D-53117 Bonn, Germany\\
$^3$Institut f\"ur Festk\"orperforschung, Forschungszentrum
J\"ulich, D-52425 J\"ulich, Germany  }

\date{\today }

\begin{abstract}
We consider the asymmetric simple exclusion process (ASEP) on a semi-infinite
chain which is coupled at the end to a reservoir with a particle density that
changes periodically in time. It is shown that the density profile assumes a
time-periodic sawtooth-like shape. This shape does not depend on initial
conditions and is found analytically in the hydrodynamic limit. In a finite
system, the stationary state is shown to be governed by effective boundary
densities and the extremal flux principle. Effective boundary densities are
determined numerically via Monte Carlo simulations and compared with those
given by mean field approach and numerical integration of the hydrodynamic
limit equation which is the Burgers equation. Our results extend
straightforwardly beyond the ASEP to a wide class of driven diffusive systems
with one conserved particle species.

\end{abstract}
\maketitle

%\pacs{87.14.Ee, 87.15.Aa, 87.15.Vv}

\section{Introduction}

Systems of driven diffusing particles attract attention because, despite their
relative simplicity, they embrace a whole range of critical phenomena far from
thermal equilibrium \cite{Schm95,Priv97,Schu00}. One of the remarkable
features of these systems is the appearance of phase transitions induced by
spatial boundaries of an open system which exchanges particles with external
reservoirs \cite{Krug91,Popk99}. A classical model where this can be studied
in great detail is the so-called Asymmetric Simple Exclusion Process (ASEP)
with open boundaries. This model describes the single-file random motion of
particles with hard core exclusion and drift inside a finite system at the
ends of which particles can be extracted or injected with some rates. This
model was first introduced for describing the kinetics of protein synthesis
\cite{MacD68,Schu97} and has since then been generalized in many ways for
describing the motion of various kinds of molecular motors
\cite{Parm03,Popk03,Nish05}. Due to its conceptual simplicity it also plays a
fundamental role in traffic flow theory \cite{Chow00,Helb01} and many other
settings where driven diffusion of interacting particles plays a role.

By now, the dynamics of the ASEP (as well as its stationary bulk behavior) is
rather well understood. For our purposes we note that the exact stationary
distribution has been determined analytically \cite{Schu93,Derr93}, while the
coarse-grained dynamics of shocks and localized excitations in the evolution
of the particle density can be understood using hydrodynamic limit equations
\cite{Spoh90,Kipn99}. The latter provides a full description of the evolution
of the local density under Eulerian scaling. It has been shown rigorously
\cite{Baha07} to be given by the famous Burgers equation used for the
description of the dynamics of shocks in dissipative systems \cite{Whitham}.

The vast body of knowledge about the ASEP has been obtained for
time-homogeneous conditions where the boundary rates are kept constant in
time. In contrast, very little is known when the environment of this open
system changes non-adiabatically in time on scales that are comparable to the
macroscopic Eulerian hydrodynamic regime. This has to be modelled by
time-dependent boundary rates which, to our knowledge, has not yet been
attempted for the ASEP with open boundaries. It is the purpose of this work to
report simulation results for a natural time-periodic setting and to analyze
these data in the framework of the hydrodynamic theory.

The paper is organized as follows. In Sec. \ref{Sec2} we define the model and
present our simulation data. The data for the Eulerian low-frequency regime of
a semi-infinite system are then analyzed and explained by extending
hydrodynamic theory to incorporate time-dependent boundary conditions
(\ref{Sec3}). This analysis allows us to predict the phase diagram of an open
finite system with two boundaries. This prediction and its numerical
verification is given in Sec. \ref{Sec4}. We end with a summary of our results
and some conclusions (\ref{Sec5}).

\section{ASEP model on a semiline with time periodic boundary}

\label{Sec2}

We consider the ASEP defined on a semi-infinite chain $\{k\in\mathbb{Z}%
,,k\leq0\}$ with a right boundary site $k=0$. Each site of the chain is either
occupied by one particle or empty. We denote the local occupation number by
$n_{k}\in\{0,1\}$. Particles attempt to jump to the right or to the left
neighboring site after an exponentially distributed random time with parameter
$p+q$, normalized as $p+q=1$. The rate at which a particle attempts to hop to
the right (left) is $p$ ($q$). If the target site is empty, the attempt is
successful and the particles moves. Otherwise, it does not jump (hardcore
exclusion rule). At the boundary site $k=0$ a particle can be extracted with
the rate $\beta$ (if the boundary site is occupied) or be injected into it (if
the boundary site is empty) with rate $\delta$. We choose $\beta=p(1-\rho
_{R})\;$, $\delta=q\rho_{R}$ so that the boundary may be thought of as being
coupled to a reservoir of density $\rho_{R}$ \cite{Popk04a}.
\begin{figure}[ptb]
\begin{center}
\includegraphics[
height=2.3151in, width=2.9014in ] {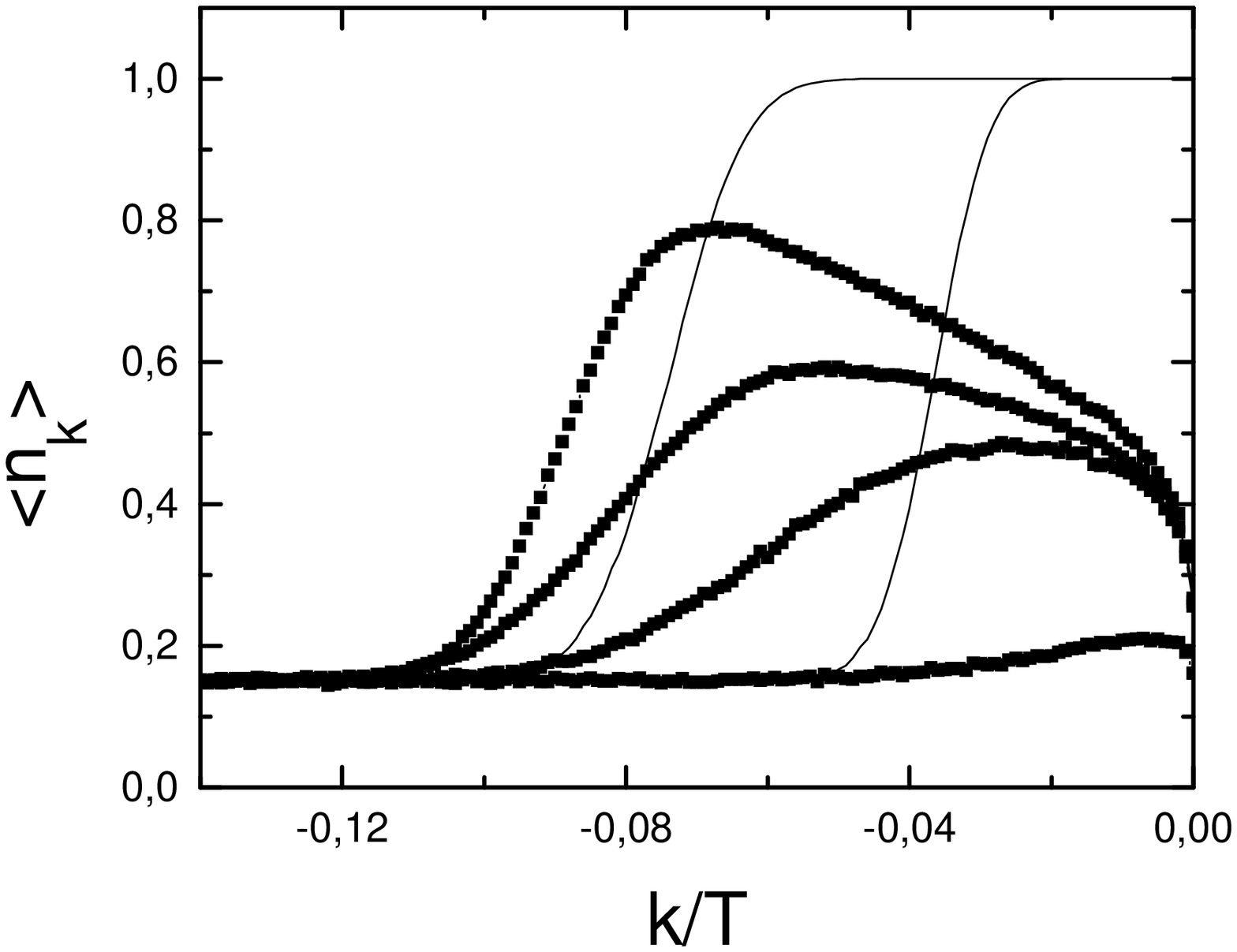}
\end{center}
\caption{Snapshots of density profiles averaged over 3*10$^{4}$ histories,
during red light $t=0.75T,t=T$(lines) and during the green light
$t=1.1T,1.2T,1.3T,1.5T$ (points). Parameters: $T=1000$. Initial state is a
homogeneous one with the density $\langle n_{k}(0)\rangle=0.15$. }%
\label{Fig_LD015}%
\end{figure}We consider the boundary reservoir density $\rho_{R}$ to be a
periodic function of time with frequency $\nu$, switching between the values
$\rho_{R}=0$ and $\rho_{R}=1$ according to
\begin{equation}
\rho_{R}\left(  t\right)  =\frac{1}{2}\left(  1+sign[\sin(2\pi\nu t)]\right)
. \label{BC_heaviside}%
\end{equation}
In traffic flow problems such a boundary condition models a traffic light with
$\rho_{R}(t)=0$ during the "green light" half-periods $t\subset\lbrack
0,\frac{T}{2}],[T,\frac{3T}{2}],...$, and $\rho_{R}(t)=1$ during "red light"
half-periods $\tau_{red}\subset\lbrack\frac{T}{2},T],[\frac{3T}{2},2T],...$ ,
where $T=1/\nu$. In analogy with this we refer to Eq.(\ref{BC_heaviside}) as
traffic light boundary condition. For a study of such a switching in a related
system, see \cite{Broc01}.

\begin{figure}[ptb]
\begin{center}
\includegraphics[
height=2.1151in, width=2.7014in ] {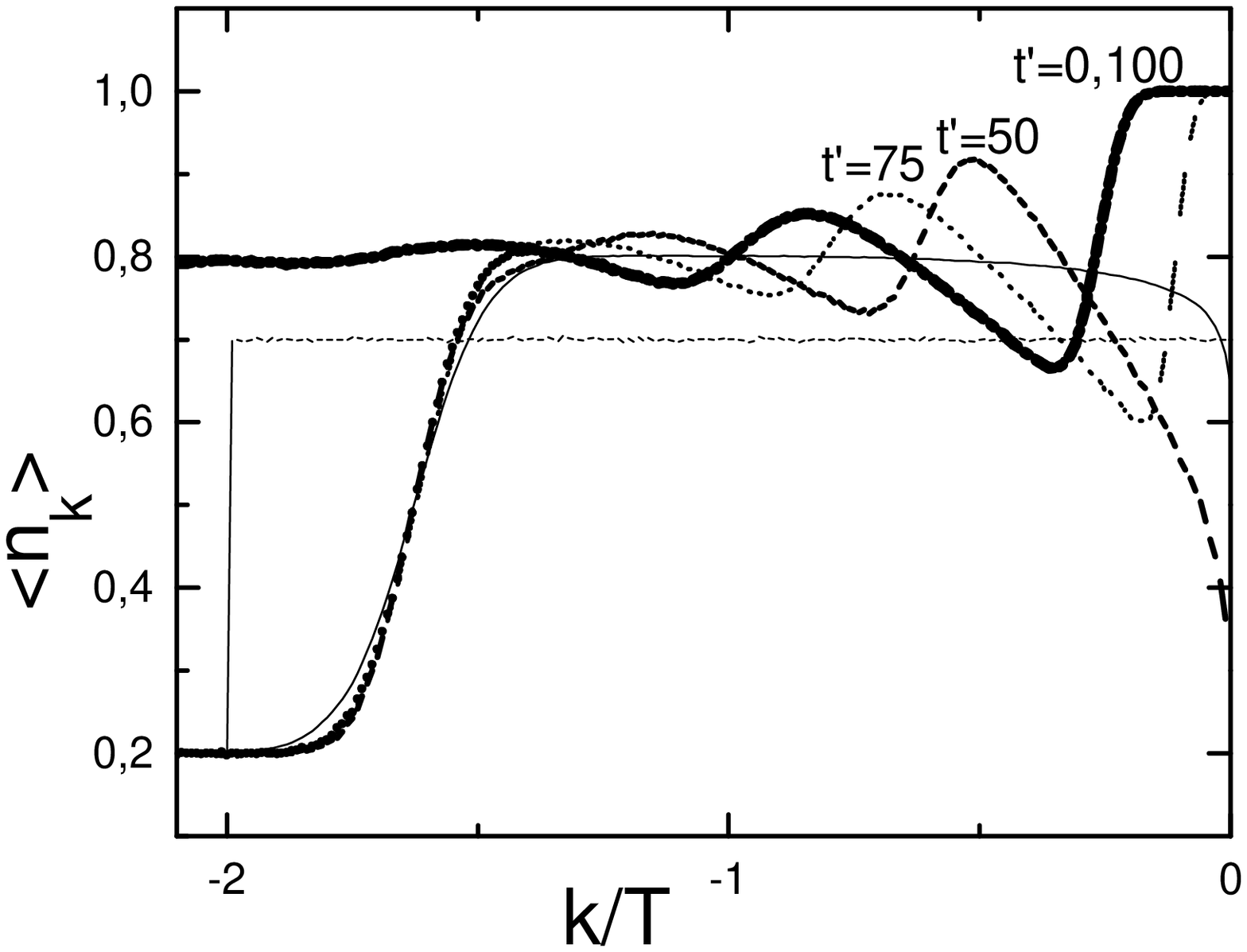}
\end{center}
\caption{Snapshots of density profiles at times $t^{\prime}=t-t_{0}%
=0,50,75,100$, at the coexistence line between LD and HD states $j_{in}%
=j_{out}$. Parameters: $T=100$. The initial configuration of the system is a
domain wall $0.2,0.7$ positioned at $k/T=-2$ ( dotted piecewise straight
line). The system was equilibrated for $t_{0}=5T=500$ before the measurements,
and the averaging over 10$^{5}$ histories was done. Thick line shows the
density profile after $t=800$ in the system prepared initially in HD state
(taken from Fig.\ref{Fig_stro}). Thin line shows the density profile averaged
over several periods and many histories after initial transient period of $5T$
and it looks (apart from boundary layer) like unbiased domain wall
$(0.2,0.8=\rho_{R}^{eff}(\nu)$ in TASEP with \textit{constant} boundary rates
}%
\label{Fig_LDHDTransition}%
\end{figure}

\begin{figure}[ptb]
\begin{center}
\includegraphics[
height=2.4017in, width=3.3155in ]{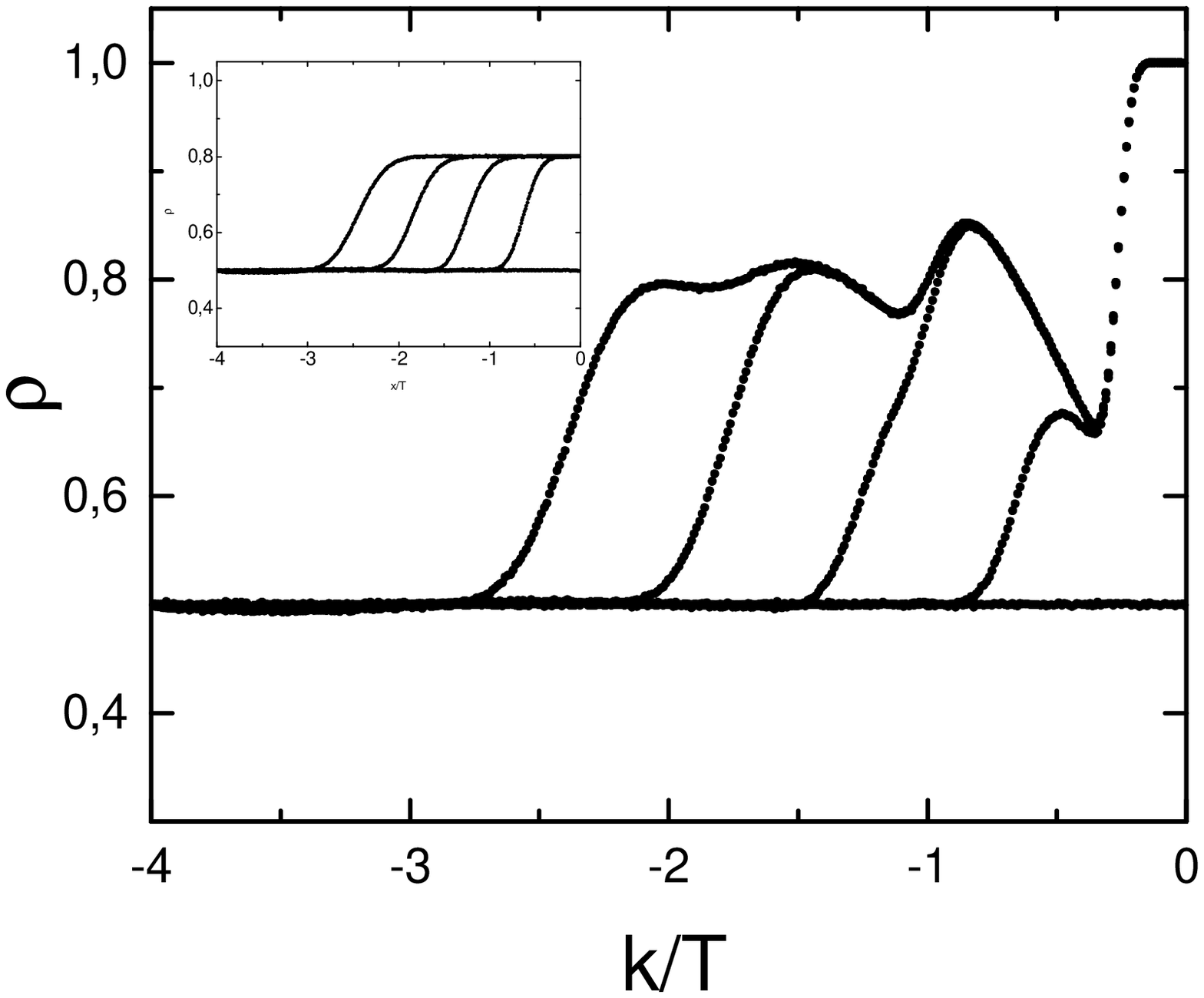}
\end{center}
\caption{Snapshots of averaged density profiles $\rho(x,t)$ at $t=0$
(homogeneous state with the density $\langle n_{k}\rangle=0.5$) and then at
equal intervals of time (equal to two periods $T$) $t_{i}=200,400,600,800$.
Parameters: $\nu=T^{-1}=0.01$, system size $400$, left boundary density $0.5$,
averaging is done over 3*10$^{5}$ histories. Inset shows the snapshots of
density profiles $\langle n_{k}\rangle$ at times $t=t_{i}$ in case when
instead of traffic light conditions, time-independent boundary conditions are
applied with the same effective right boundary density $\rho_{R}=\rho
_{R}^{eff}(\nu)\approx0.8$. }%
\label{Fig_stro}%
\end{figure}

\begin{figure}[ptb]
\begin{center}
\includegraphics[
height=2.4017in, width=3.3155in ]{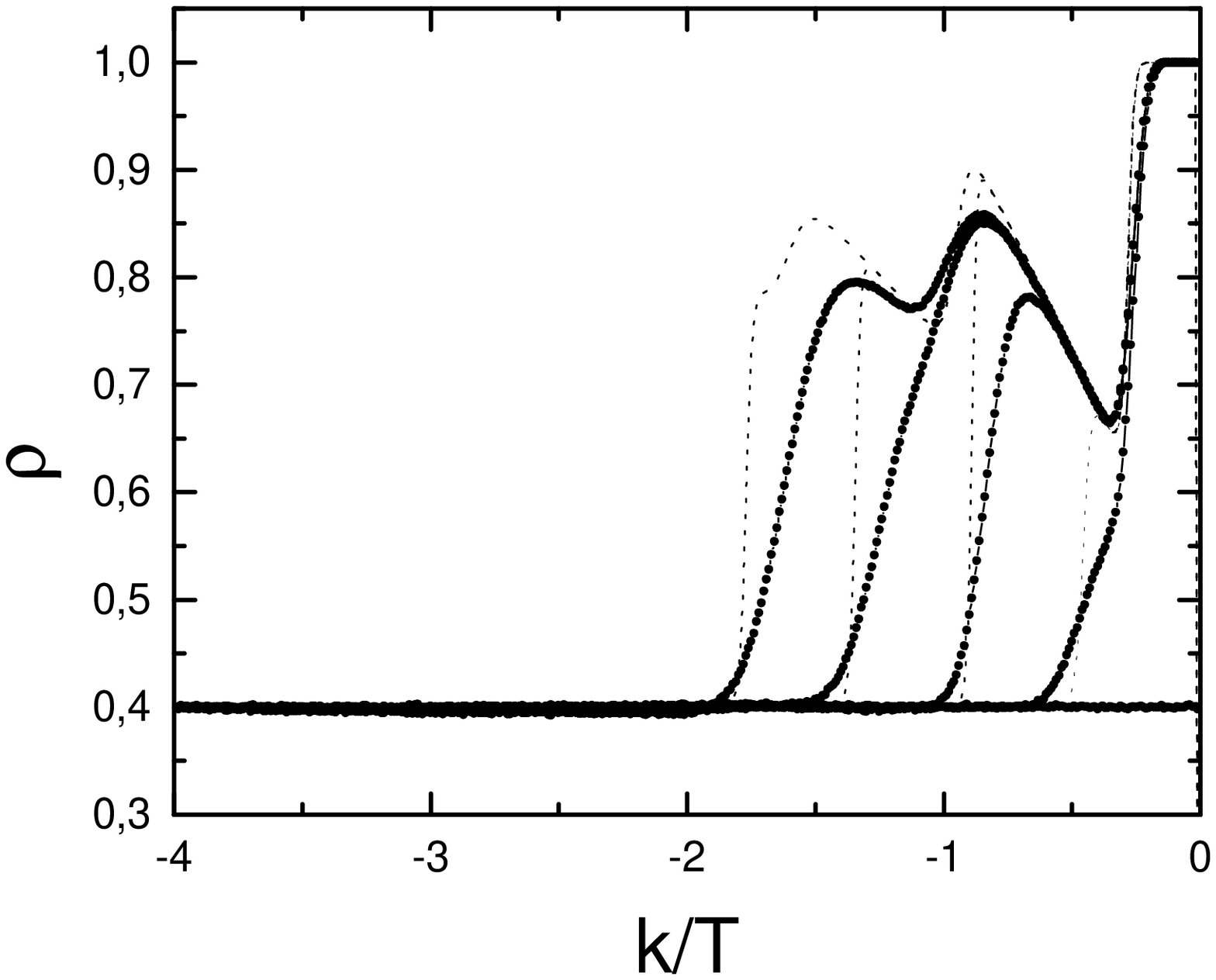}
\end{center}
\caption{ Same as in Fig.\ref{Fig_stro} but for an initial homogeneous state
with density $\rho(x,0)=0.4$. Dotted lines show the result on numerical
integration of mean field equations (\ref{MeanfieldEq}) with $p=1$. }%
\label{Fig_stro04}%
\end{figure}

To investigate the effects of this boundary condition on the dynamics of the
ASEP on a semiline we have carried out Monte-Carlo simulations of the model
for various frequencies $\nu$. We have concentrated mainly on the case of a
totally asymmetric simple exclusion process (TASEP), $q=0,p=1$, and focused
our interest on stationary behavior in the sense, that all macroscopic
quantities behave periodically in time with period $T$. The initial state was
prepared in an ensemble of particles randomly distributed with the density
$\lambda$, which is stationary in the infinite system. In presence of
periodically varying boundary rates, rather complicated dynamics is observed.
For small ingoing fluxes $j_{in}=\lambda\left(  1-\lambda\right)  $, the
system develops jam at the boundary during the red light periods, which is
dissolving completely during the green light periods, Fig.\ref{Fig_LD015}.
However, if $\lambda$ (and consequently, the inflow flux) exceeds some
critical value $\lambda>\lambda_{c}(\nu)$, the jam at the boundary is not
dissolved completely, but starts to propagate inside the system. The amplitude
of the shock front is not constant, it increases and drops during each
red-green period, see Fig.\ref{Fig_LDHDTransition}. Consequently, shock front
is not propagating steadily, but its velocity changes, and in particular it
may advance and retract during each red-green period. The net shock advance after a period $T$ is determined by the mass
conservation, i.e. difference between the ingoing $j_{in}T$ and outgoing
fluxes of particles $j_{out}T$, see Figs.\ref{Fig_stro},\ref{Fig_stro04}%
.\ While $j_{in}$ is a control parameter $j_{in}=\lambda\left(  1-\lambda
\right)  $, $j_{out}$ is not and it is measured as the time-averaged particle
flux through the boundary. $j_{out}$ depends on the frequency of traffic light
$\nu$ and it is associated with the effective right boundary density $\rho
_{R}^{eff}(\nu)$ through the TASEP current-density relation $j_{out}=\rho
_{R}^{eff}(\nu)\left(  1-\rho_{R}^{eff}(\nu)\right)  $ as discussed below.
%
%Fig. 3,4 splittata
%%%%%%%%%%%%%%%%%%%%%%%%%%%%%%%%%%%%%%%%%%%%%%%%%%%%%%%%%%%%%%%%%%%%%%
If ingoing and outgoing fluxes are equal, the shock only
"breathes" around its initial position see Fig.\ref{Fig_LDHDTransition}. The
density profile behind the shock front develops approximately equidistant
sawteeth-like structures with the decaying amplitudes, see Fig.\ref{Fig_stro}.
The sawteeth profile is changing with time, but it regains its shape after
each complete period, so that the shape depends only on the phase
$\gamma\subset\lbrack0,T]$. Apart from this phase dependence, illustrated e.g.
on Fig.\ref{Fig_LDHDTransition}, the sawteeth structure behind the shock front
depends on frequency $\nu$ (the rescaled sawteeth become sharper with
decreasing $\nu$ ) but not on $\lambda$, see Figs.\ref{Fig_LDHDTransition}%
,\ref{Fig_stro},\ref{Fig_stro04}. \begin{figure}[ptb]
\begin{center}
\includegraphics[
height=2.4238in, width=2.8536in ] {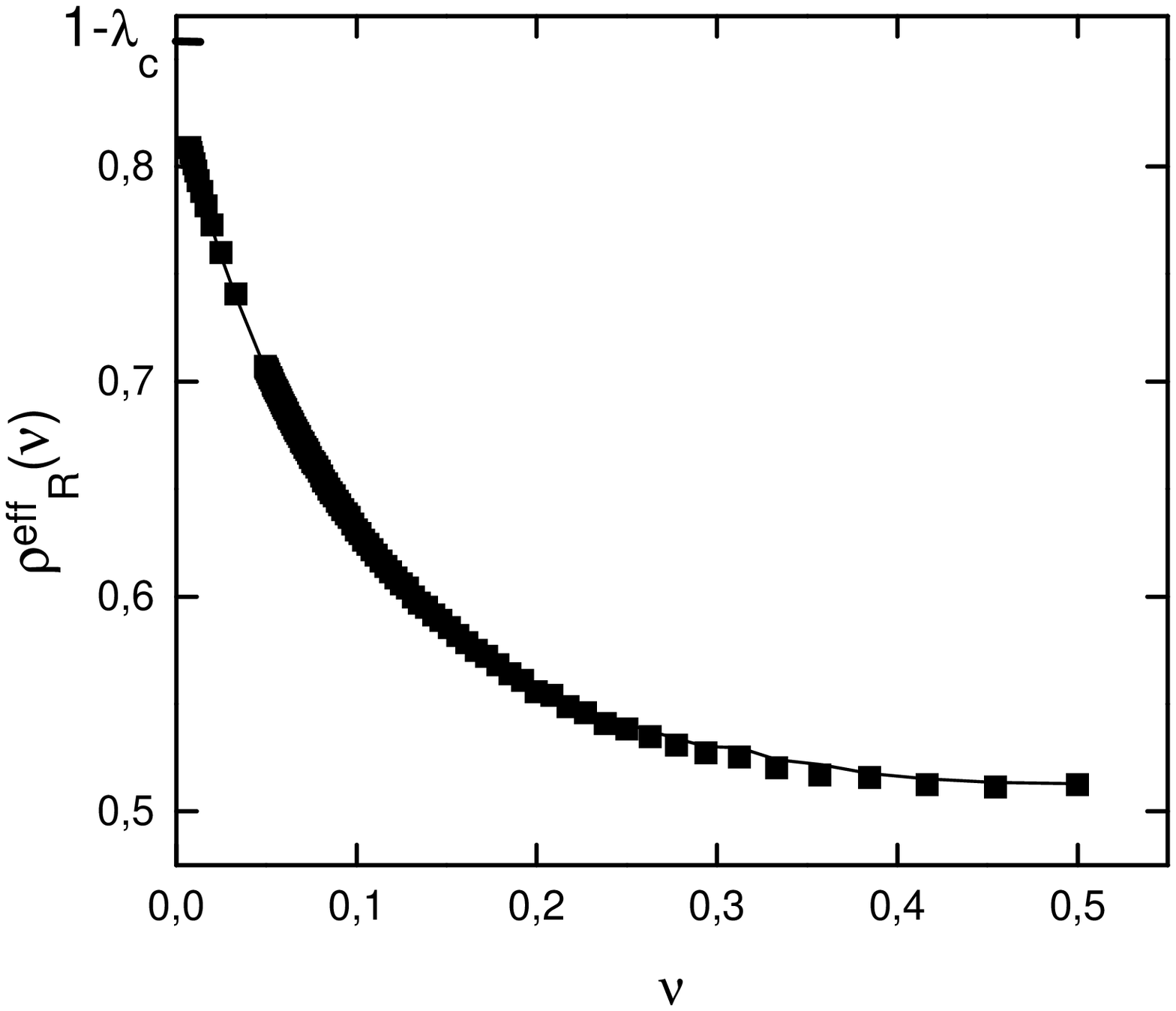}
\end{center}
\caption{The averaged stationary density (for parameters chosen equal to
effective right boundary density $\rho_{R}^{eff}$) in the TASEP\ versus the
frequency of "traffic light" switch at the right boundary, $\nu=T^{-1},$ from
Monte Carlo simulations. The parameters are: system size $400$, the left
boundary density is higher than $1/2$, and the averaging is done over $30$
histories and over $2\ast10^{5}$ Monte Carlo steps, after the equilibration.
Points show the results of density averages, and the broken line reports
estimates from stationary flux measurements. }%
\label{Fig_roplus}%
\end{figure}The particle density at some distance from the boundary approaches
a constant value, $\rho_{R}^{eff}(\nu)$. This value can be identified with an
effective right boundary density in the following sense: the particle system
(ASEP) behaves as if it was joined at the origin with the reservoir of
particles with the density $\rho_{R}=\rho_{R}^{eff}(\nu)$. As a function of
frequency the averaged stationary flux through the boundary changes
monotonically, which implies a monotonic change of $\rho_{R}^{eff}(\nu)$ . In
Fig.\ref{Fig_roplus} we report the numerical results for $\rho_{R}^{eff}(\nu)$
from the Monte-Carlo simulations.

\section{Hydrodynamic theory and sawtooth structure}

\label{Sec3}

\subsection{Limiting cases}

In order to understand the origin of these observations we first argue that if
$\lambda\geq1/2$ then the right boundary reservoir density $\rho_{R}%
^{eff}\left(  \nu\right)  $ must indeed be equal or larger than $1/2$ for any
value of $\nu$, as observed: During the "green light" periods $\beta=1$ at
most 1 particle per time unit can exit, while during the "red light" periods
$\beta=0$ no particles can exit. For frequencies large compared to the mean
attempt rate for particle jumps (which is 1), a particle at the boundary site
``sees'' the reservoir with equal probability empty or fully occupied,
\textit{irrespectively of how long it has already stayed at the boundary
site}. Hence the system behaves like a time-homogeneous system where at each
time a particle can exit with the effective rate $\beta^{eff}=1/2$, which
corresponds to $\rho_{R}^{eff}=1-\beta^{eff}=1/2$ in TASEP. So we have
$\lim_{\nu\rightarrow\infty}\rho_{R}^{eff}\left(  \nu\right)  =1/2$ which we
expect to be a good approximation for all frequencies $\nu\gg1$ much larger
than the jump attempt rate.

On the other hand, in the static case of zero frequency the system relaxes
into the high density phase by a back-moving shock if the traffic light cycle
starts with $\beta=0$. This leaves the system with a bulk density $\rho=1$. If
started with a green traffic light, $\beta=1$, the system reaches the
maximal-current phase with bulk density $\rho=1/2$ \cite{Kolo98}. Continuity
in frequency then gives $\rho_{R}^{eff}\left(  \nu\right)  \geq1/2$ for all
frequencies $\nu$ provided that $\lambda\geq1/2$ \footnote{For a driven
particle model with arbitrary current-density relation $j\left(  \rho\right)
$ which has a single maximum $\max j\left(  \rho\right)  =j\left(  \rho^{\ast
}\right)  =j_{\max}$, it is expected that the stationary flux through the
boundary equals the maximal possible flux $j_{\max}$, i.e. $\lim
_{\nu\rightarrow\infty}\rho_{R}^{eff}\left(  \nu\right)  =\rho^{\ast}$.}. If
the system is initially in the low density phase, then in the high frequency
limit one has again an effective right boundary density of 1/2, leaving the
system in the low density phase with a bulk density $\lambda$. In the
zero-frequency case the limiting behavior depends again on how the period
starts. If $\beta=0$ (red traffic light), the system fills up as described
above and $\rho=1$. On the other hand, for $\beta=1$, the system remains in
the low density phase with a bulk density $\lambda$. \qquad

Notice that neither of these zero-frequency behaviors represents the
zero-frequency limit $\nu\rightarrow0$ shown in Fig.\ref{Fig_roplus}. For this
limit we demonstrate in Sec. \ref{Sec4} that
\begin{equation}
\lim_{\nu\rightarrow0}j\left(  \rho_{R}^{eff}\left(  \nu\right)  \right)
=\frac{j_{\max}}{2}\text{ \ \ and }\rho_{R}^{eff}\left(  \nu\right)
>\,\rho^{\ast}. \label{lim_smallnu}%
\end{equation}
This result is based on hydrodynamic limit arguments which are the subject of
the following subsections.

\subsection{Hydrodynamic limit and mean field description}

The most basic question to be asked about the dynamics of an interacting
particle system is its large-scale behavior, i.e., how macroscopic equations
of motion arise from its microscopic dynamics. By suitable coarse-graining of
space and time the law of large numbers usually guarantees that stochastic
variables, in the present case the particle number in some interval (which
under scaling becomes infinite on microscopic scale, but still infinitesimal
on macroscopic scale), turn into mean values whose temporal evolution satisfy
some deterministic evolution equation in rescaled macroscopic time. Moreover,
on macroscopic time scales the system is locally stationary, i.e., all fast
variables not captured in the evolution equation are locally stationary. This
fact determines the precise form of the macroscopic equation, provided the
stationary distributions are known.

In the case of a conserved quantity the evolution equation is a conservative
pde of the form
\begin{equation}
{\frac{\partial\rho}{\partial\tau}}+{\frac{\partial\left(  j\left(
\rho\right)  \right)  }{\partial x}=0} \label{conservation law}%
\end{equation}
where $x$ is the rescaled space variable and $\tau$ is rescaled macroscopic
time \cite{Spoh91,Kipn99}. The quantity $j\left(  \rho\right)  $ is the
particle current which on the macroscopic time scales for which
(\ref{conservation law}) is valid takes its locally stationary value at
density $\rho(x,\tau)$. In the case of the ASEP one has $j(\rho) = (p-q)
\rho(1-\rho)$. Generally, for driven particle systems with a finite
macroscopic current $j(\rho)$ the appropriate hydrodynamic scale is the Euler
scale $\tau=ta$ $x=ka$, where $a\to0$ is the lattice constant. For the ASEP
investigated here we get the inviscid Burgers equation
\begin{equation}
{\frac{\partial\rho}{\partial\tau}}+(p-q){\frac{\partial\left(  \rho
(1-\rho)\right)  }{\partial x}=0} \label{InviscidBurgersEq}%
\end{equation}
on the semiline $x\leq0$ with traffic light boundary conditions on the right
boundary $x=0$. For the TASEP we set $p=1$ and $q=0$.

Notice that in the present case of periodic boundary driving also the period
has to be rescaled. The boundary conditions become $\rho\left(  0,\tau\right)
=\frac{1}{2}\left(  1+sign[\sin(2\pi\Omega t)]\right)  $ where $\Omega=\nu/a$
is the rescaled frequency. I.e. the boundary stays open ($\rho(0,\tau
_{green})=0$ ) during green light half-periods $\tau_{green}\subset
\lbrack0,\frac{T}{2}],[T,\frac{3T}{2}],...$ and closed $\rho(0,\tau_{red})=1$
during the "red light" half-periods $\tau_{red}\subset\lbrack\frac{T}%
{2},T],[\frac{3T}{2},2T],...$. Here we denote by $T$ the complete period
$T=1/\Omega$.

The inviscid Burgers equation is the zero-viscosity limit of the viscous
Burgers equation
\begin{equation}
{\frac{\partial\rho}{\partial\tau}}+{\frac{\partial\left(  \rho(1-\rho
)\right)  }{\partial x}}=D{\frac{\partial^{2}\rho}{\partial x^{2}},}
\label{BurgersEquation}%
\end{equation}
which can be solved in explicit form by a Hopf-Cole transformation for fairly
general boundary conditions \cite{Calogero}. For traffic light boundary
conditions, however, such a solution is difficult to obtain and we solve the
coarse-grained time evolution of the ASEP by numerical integration. To this
end we note that the exact microscopic operator equations of motion for the
expected particle number $n_{k}$ on site $k$ read
\begin{align*}
\frac{\partial}{\partial t}\langle n_{k}\rangle &  =p\langle n_{k-1}\left(
1-n_{k}\right)  \rangle-q\langle n_{k}\left(  1-n_{k-1}\right)  \rangle\\
&  -p\langle n_{k}\left(  1-n_{k+1}\right)  \rangle+q\langle n_{k+1}\left(
1-n_{k}\right)  \rangle
\end{align*}
In this equation a one-point function (the expected density $\langle n_{k}
\rangle$) is coupled to two-point functions (on the r.h.s. of the equation),
i.e. the equation is not closed. Writing down an exact equation for the
two-point functions introduces three-point functions and so on. This infinite
hierarchy of equations is not directly tractable and some closure scheme must
be employed for further analysis.

In the mean field approximation for the ASEP, we neglect the correlations and
approximate $\langle n_{k}n_{k+1}\rangle=\langle n_{k}\rangle\langle
n_{k+1}\rangle=\rho_{k}\rho_{k+1}$ etc. where $\rho_{k}=\langle n_{k}\rangle$
is an average particle density at site $k$. Using this approach for the exact
microscopic evolution equation we obtain after some algebra
\begin{align}
\frac{\partial}{\partial t}\rho_{k}  &  =-\left(  p-q\right)  \left[  \left(
1-2\rho_{k}\right)  \frac{\rho_{k+1}-\rho_{k-1}}{2}\right] \nonumber\\
&  +\left(  p+q\right)  \left[  \frac{\rho_{k-1}+\rho_{k+1}-2\rho_{k}}%
{2}\right]  . \label{MeanfieldEq}%
\end{align}
These equations are complemented with the traffic-light boundary conditions
$\rho_{0}\left(  t\right)  =\rho_{R}\left(  t\right)  $ in (\ref{BC_heaviside}).

Some comments are in order. Firstly, in the continuum limit one substitutes
$ka=x$, with $a\ll1$ being the lattice constant, e.g. $\rho_{k+1}\left(
t\right)  \rightarrow\rho\left(  x+a,t\right)  $. Taylor-expanding
(\ref{MeanfieldEq}), and using ($p+q=1$) we obtain, after rescaling time as
$\tau=ta$, the Burgers equation
\begin{equation}
{\frac{\partial\rho}{\partial\tau}}+\left(  p-q\right)  {\frac{\partial\left(
\rho(1-\rho)\right)  }{\partial x}}=\frac{a}{2}{\frac{\partial^{2}\rho
}{\partial x^{2}}} \label{BurgersEqD}%
\end{equation}
Therefore, the mean field equation (\ref{MeanfieldEq}) can be viewed as a
discretization of the viscous Burgers equation (\ref{BurgersEquation}), with a
constant discretization step $a,$ and $D=a^{2}/2$. In fact, for numerical
integration of the Burgers equation (\ref{BurgersEquation}) with the diffusion
coefficient $D$, we shall be using the discretization scheme
(\ref{MeanfieldEq}) with $p=1,q=0$, discretization step $a=1$ and the
coefficient $2D$, instead of $\left(  p+q\right)  $ in front of discrete
second derivative.

\begin{figure}[ptb]
\begin{center}
\includegraphics[
height=2.2935in, width=2.6645in ]{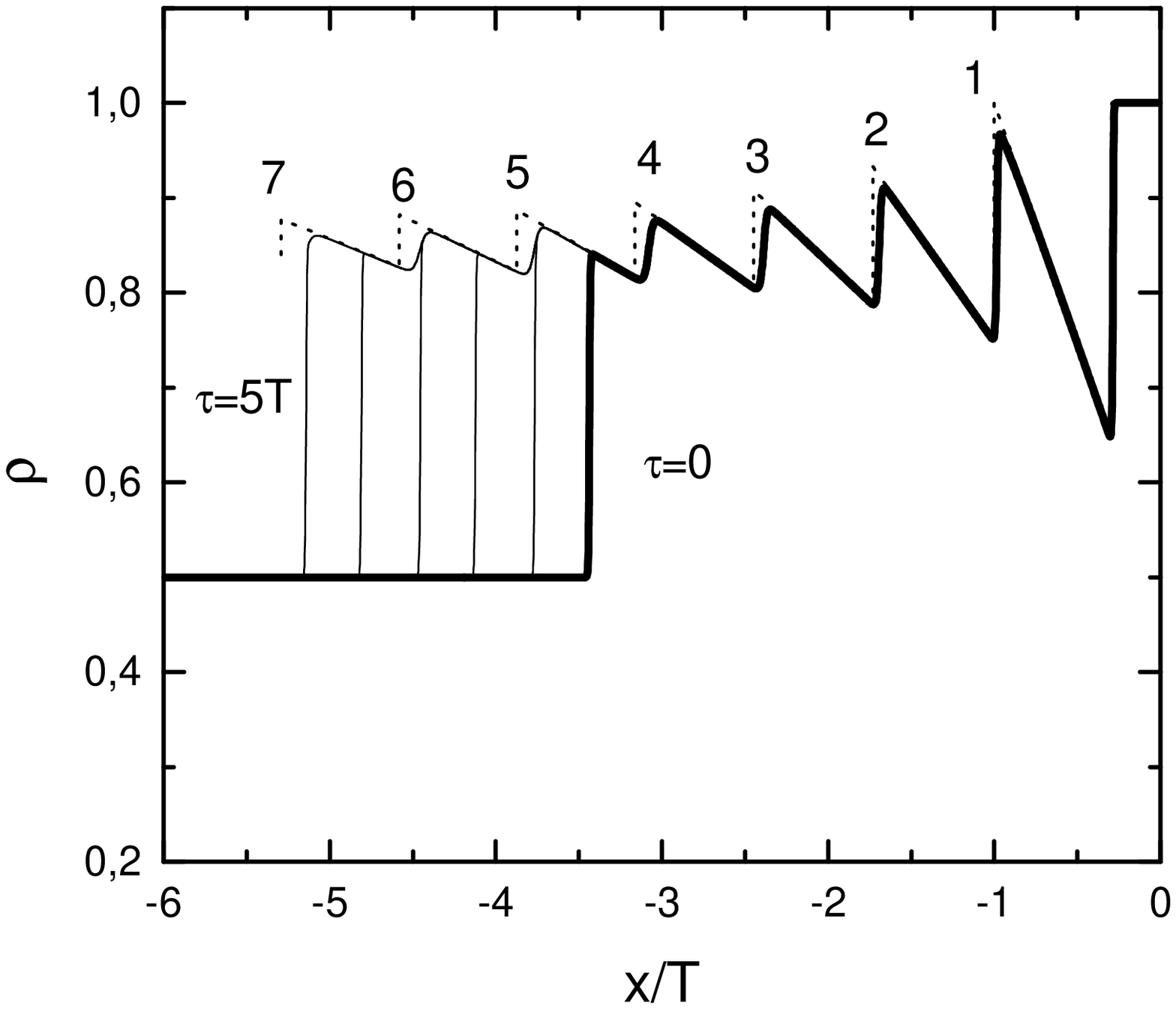}
\end{center}
\caption{Results of numerical integration of mean field equations
(\ref{MeanfieldEq}) for system size $N=3000,\nu=T^{-1}=0.002,$ with initial
condition $\rho(x)=0.5$ after $10$ full periods (thick curve). Thin curves
show the subsequent density profiles $\rho(x,\tau)$ after $1,2,3,4$ and $5$
full periods. The broken line shows the curve $\Gamma$ (\ref{h_k}%
-\ref{alpha_k}) exact in the limit $\nu\rightarrow0$. }%
\label{Fig_DentiMF}%
\end{figure}

Secondly, for the case of weak hopping asymmetry $\lim_{a\rightarrow0}a\left(
p-q\right)  =1$ one can obtain Eq. (\ref{BurgersEquation}) from
(\ref{MeanfieldEq}) by \textit{diffusive} rescaling $\tau=ta^{2}/D$,
$x^{\prime}=xa^{2}/D$ in the hydrodynamic limit $a\rightarrow0$. In this case,
the density does not evolve into shocks. A stationary travelling wave solution
is a hyperbolic tangent with a step width proportional to the square root of
the viscosity. Such smoothening of a shock may also be expected from the
integration of the discrete mean field equation \ref{MeanfieldEq}. In Fig.
\ref{Fig_DentiMF} we show the results obtained from numerical integration of
the mean field equations (\ref{MeanfieldEq}) when a traffic light boundary is
present at the origin. We see that the density profiles $\rho(x,\tau)$ display
a sawtooth structure which resembles the one observed for ASEP (see Fig.
\ref{Fig_stro}). 

As a warning to readers not familiar with hydrodynamic scaling, we remark that
the agreement between the mean field equation for the ASEP in the continuum
limit and the rigorously derived Burgers equation is purely coincidental. It
arises from the fact that for the ASEP the stationary distribution has no
correlations and hence the mean field equations become exact. In general
lattice gases, e.g. in the KLS-model \cite{Popk99}, one has correlations and a
simple-minded mean field approach neglecting all correlations would produce a
macroscopic equation that is in general not even qualitatively correct.
\begin{figure}[ptb]
\begin{center}
\includegraphics[
height=2.4872in, width=3.0225in ] {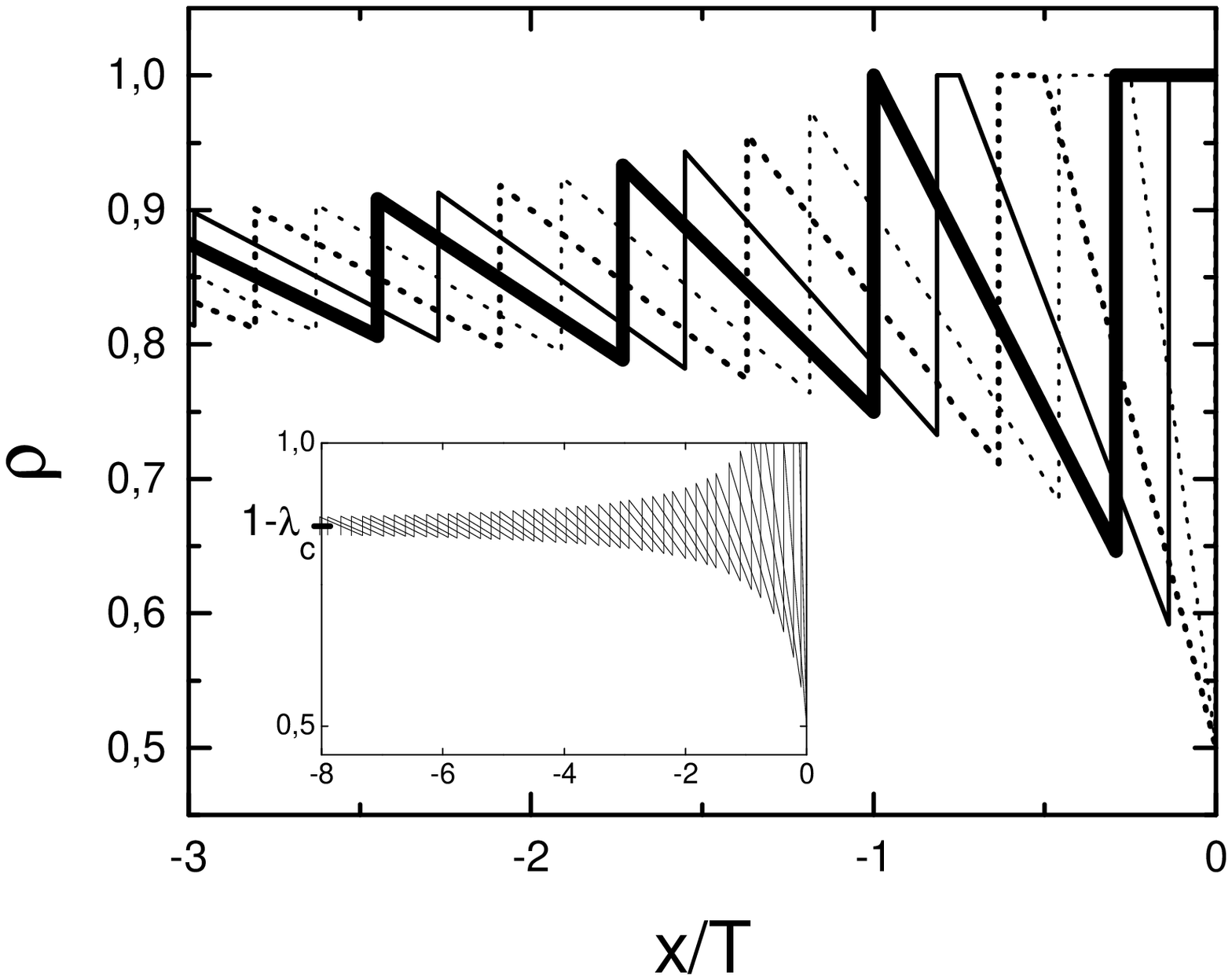}
\end{center}
\caption{The universal curve $\Gamma$ (solid curve) at the beginning of the
green light period $\tau=0,T,2T,...$(thick curve) and for intermediate $\tau$
values ${\tau/T}=1/4,1/2,3/4$ (broken, thick broken, thin curves,
respectively). Inset shows $\Gamma$ snapshots over larger scale. Note that
microscopically the snapshots of average density profiles during the green
light period $\rho_{R}=0$ do not have discontinuity at the first site, but a
boundary layer involving many sites, interpolating between $\rho=0.5$ and
$\rho_{R}=0$. However, the $\Gamma$ snapshots show the limiting density
profiles, rescaled by the period $T$, in the limit when $T\rightarrow\infty$
($\nu=+0$). The boundary layer vanishes in this limit leading to a
discontinuity at $x=0$.}%
\label{Fig_Denti_Envelope}%
\end{figure}

\subsection{Stationary sawtooth structure}

The time-periodic sawtooth solution shown in Fig.\ref{Fig_DentiMF} is a shock
analog for the case of periodically changing boundary conditions which we now
describe in detail. To this regard, we take the analytically tractable Burgers
case (\ref{InviscidBurgersEq}) as a concrete example for explicit computation.
The results for stationary (periodically repeating ) solutions obtained in
this case are expected to be valid also for generic conservation law equation
with convex $j(\rho)$.

Looking at snapshots of density profiles $\rho(x,t+mT)$ at times differing by
multiples $mT$ of a period, one notes that they fill some universal curve
$\Gamma$, which has a characteristic sawtooth shape, see Fig.\ref{Fig_DentiMF}%
. The curve $\Gamma_{\gamma}(x,\tau)=\Gamma_{\gamma}(x,\tau+T)$ "breathes" and
returns to its original form after a full time period $T$, see
Fig.\ref{Fig_Denti_Envelope}. The index $\gamma$ denotes a phase $0\leq
\gamma<T$ at which the snapshot of $\Gamma$ is taken, with respect to the
beginning of a green light interval. In the following we shall set $\gamma=0$
and omit $\gamma$ for brevity of notation.

\begin{figure}[ptb]
\begin{center}
\includegraphics[
height=2.4872in, width=3.0225in ] {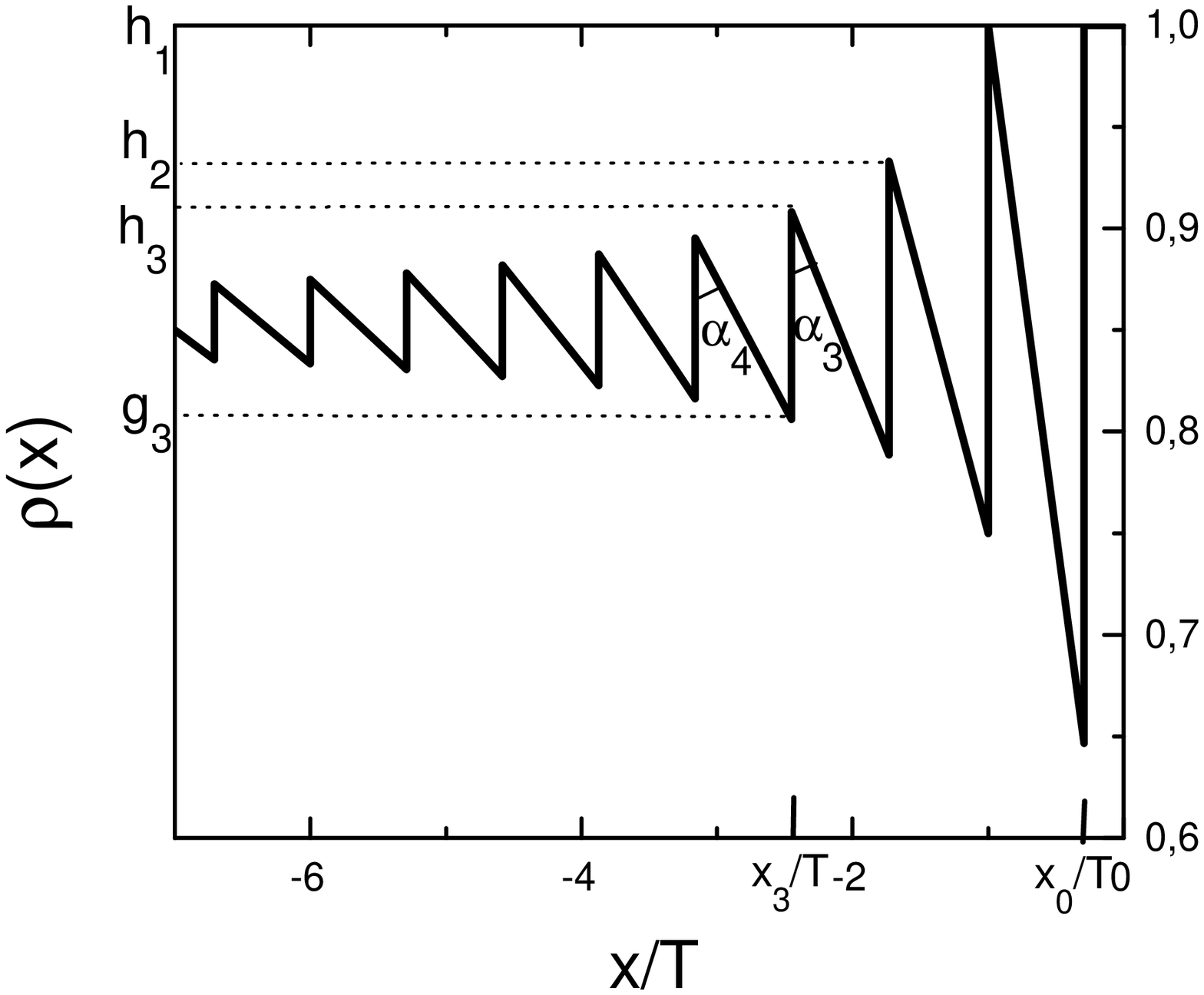}
\end{center}
\caption{ The universal curve $\Gamma$ (solid curve) at the beginning of the
green light period and variables $g_{k},h_{k}, \alpha_{k}$ used in the
analysis (\ref{h_k}-\ref{alpha_k}). }%
\label{Fig_DentiOrigin}%
\end{figure}

Firstly, we describe the curve $\Gamma$ and then prove its periodicity in
time. The curve consists of infinite number of sawteeth with heights
decreasing away from the boundary. We denote the height, the base and the
coordinate of $k$-th sawtooth at the beginning of green light periods by
$h_{k}^{0},g_{k}^{0}$ and $x_{k}^{0}$, respectively, and the sawtooth angle by
$\alpha_{k}^{0}$ as shown in Fig.\ref{Fig_DentiOrigin}. Each sawtooth is
bounded by a shock discontinuity on the left and by a rarefaction wave on the
right except the sawtooth $k=0$ bounded on the right by a jam (caused by just
finished red light period). Shock discontinuities move with the velocities
given by Rankine-Hugoniot condition
\begin{equation}
v_{shock}(k)=\frac{j\left(  h_{k}\right)  -j\left(  g_{k}\right)  }%
{h_{k}-g_{k}} \label{rankine}%
\end{equation}
where $j(\rho)$ is a flux function from (\ref{conservation law}). In the
framework of stochastic driven systems $j(\rho)$ is called \ the
current-density relation or fundamental diagram. Shapes of rarefaction waves
are also determined by $j\left(  \rho\right)  $. In the following we shall
consider a specific example (\ref{InviscidBurgersEq}). However we expect that
our main results (\ref{jout=jmax/2},\ref{delta},\ref{delta_rho(x)}) are
applicable for arbitrary convex function $j\left(  \rho\right)  $.

For the specific case of inviscid Burgers equation, $h_{k}^{0},g_{k}^{0}%
,x_{k}^{0}$ and $\alpha_{k}^{0}$ for $k>0$ are given by
\begin{equation}
h_{k}^{0}=\frac{1}{2}\left(  1+\left(  1-2\lambda_{c}\right)  \sqrt{1+\frac
{1}{k}}\right)  \label{h_k}%
\end{equation}%
\begin{equation}
g_{k}^{0}=\frac{1}{2}\left(  1+\left(  1-2\lambda_{c}\right)  \sqrt{1-\frac
{1}{k+1}}\right)  \label{g_k}%
\end{equation}%
\begin{equation}
-x_{k}^{0}=kT\left(  1-2\lambda_{c}\right)  \sqrt{1+\frac{1}{k}} \label{x_k}%
\end{equation}%
\begin{equation}
\tan\left(  \alpha_{k}^{0}\right)  =2kT, \label{alpha_k}%
\end{equation}
and for $k=0$
\begin{align}
-x_{0}^{0}  &  =2\lambda_{c}T\nonumber\\
g_{0}^{0}  &  =\frac{1}{2}+\lambda_{c} \label{g_0}%
\end{align}
where $\lambda_{c}=\left(  2-\sqrt{2}\right)  /4\approx0.146$ satisfies
$\lambda_{c}\left(  1-\lambda_{c}\right)  =1/8$. For $k\gg1$, both $h_{k}%
^{0}\approx1-\lambda_{c}+\frac{1-2\lambda_{c}}{4k}$ and $g_{k}^{0}%
\approx1-\lambda_{c}-\frac{1-2\lambda_{c}}{4k}$ approach the limiting value
$1-\lambda_{c}$ which is therefore the limit of $\rho\left(  x,\tau\right)  $
as $x\rightarrow-\infty$. This value may be identified with the effective
time-independent boundary density $\rho_{R}^{eff}$. For rigorous definition of
the boundary density see \cite{Popk04a}. Here we only stress that the
underlying particle system (ASEP) behaves as if it was joined at the origin
with the reservoir of particles with the density $\rho_{R}^{eff}$.
Correspondingly, the average flux through the boundary is
\begin{equation}
\langle j_{out}\rangle=j\left(  \rho_{R}\right)  =j\left(  1-\lambda
_{c}\right)  =\frac{1}{8}=\frac{j_{\max}}{2} \label{jout_generic}%
\end{equation}

In the following we prove that $\Gamma$ is a periodic function of time
$\Gamma(\tau)=\Gamma(T+\tau) $. In this respect it is sufficient to show that
the heights and positions of all sawteeth will be the same after time $T$.
However, since all sawteeth shocks discontinuities move to the left $\partial
x_{n}\left(  \tau\right)  /\partial\tau<0$ (this follows from $h_{n}\left(
\tau\right)  >g_{n}\left(  \tau\right)  >1/2$ and (\ref{rankine})), sawteeth
cannot return to their original places. Instead, after time $T$ a sawtooth $k$
will take the place of a former sawtooth $k+1$, i.e. $h_{k}\left(  T\right)
=h_{k+1}^{0}$ , $g_{k}\left(  T\right)  =g_{k+1}^{0}$ and $x_{k}\left(
T\right)  =x_{k+1}^{0}$, for all $k$, see Fig.\ref{Fig_Denti_Envelope}. At the
boundary, the structure with a jam at position $x_{0}^{0}$ must be regenerated
after complete green light $\tau\subset\lbrack0,T/2]$ and red right
$\tau\subset\lbrack T/2,T]$ period. To proceed, note that the inviscid Burgers
equation (\ref{InviscidBurgersEq}) has two basic solution types: (a)\ between
two consecutive homogeneous states $\rho_{-}<\rho_{+}$ a shock discontinuity
is formed moving with velocity $v_{shock}=1-\rho_{-}-\rho_{+}$ and (b) between
two consecutive homogeneous states $\rho_{-}>\rho_{+}$ a rarefaction wave,
$\rho(x,\tau)=\rho(x/\tau)=1/2-x/(2\tau)$ is formed. This information is
enough to predict the evolution of sawteeth structure in
Fig.\ref{Fig_DentiOrigin}, consisting of shocks and rarefaction waves. In
particular, the shock velocity of a $k$-th sawtooth
\begin{equation}
\frac{\partial x_{k}}{\partial\tau}=1-h_{k}\left(  \tau\right)  -g_{k}\left(
\tau\right)  \label{dx_k/dt}%
\end{equation}
and
\begin{equation}
\frac{\partial}{\partial\tau}\tan\left(  \alpha_{k}\left(  \tau\right)
\right)  =2 \label{alpha_k_eq}%
\end{equation}

In the following we shall explicitly indicate time-dependent quantities e.g.
$h_{k}\left(  \tau\right)  ,\alpha_{k}\left(  \tau\right)  $ while $h_{k}%
^{0},\alpha_{k}^{0}$ will denote their initial values $h_{k}^{0}=h_{k}\left(
0\right)  $ etc.., given by (\ref{g_k})-(\ref{alpha_k}). The time interval
$0\leq\tau\leq T$ will be considered. For $k>0$ Eqs.(\ref{alpha_k_eq}%
,\ref{alpha_k}) are trivially solved: $\tan\alpha_{k}\left(  \tau\right)
=2\left(  \tau+kT\right)  $. For $h_{k}\left(  \tau\right)  ,g_{k}\left(
\tau\right)  $ we can write the equations%
\begin{equation}
\frac{-x_{k}\left(  \tau\right)  }{h_{k}\left(  \tau\right)  -\frac{1}{2}%
}=\tan\alpha_{k}\left(  \tau\right)  =2\left(  \tau+kT\right)
\label{h_k_eq_ini}%
\end{equation}%
\[
\frac{-x_{k}\left(  \tau\right)  }{g_{k}\left(  \tau\right)  -\frac{1}{2}%
}=\tan\alpha_{k+1}\left(  \tau\right)  =2\left(  \tau+\left(  k+1\right)
T\right)  .
\]
Multiplying (\ref{h_k_eq_ini}) by $h_{k}\left(  \tau\right)  -\frac{1}{2}$,
differentiating with respect to $\tau$ and using (\ref{dx_k/dt}) we obtain
\[
2\left(  \tau+kT\right)  \frac{\partial h_{k}\left(  \tau\right)  }%
{\partial\tau}+h_{k}\left(  \tau\right)  -g_{k}\left(  \tau\right)  =0.
\]
Analogously, for $g_{k}\left(  \tau\right)  $ we get an ordinary differential
equation
\[
2\left(  \tau+\left(  k+1\right)  T\right)  \frac{\partial g_{k}\left(
\tau\right)  }{\partial\tau}+g_{k}\left(  \tau\right)  -h_{k}\left(
\tau\right)  =0
\]
with the initial conditions $g_{k}\left(  0\right)  =g_{k}^{0}$,
$\ \ h_{k}\left(  0\right)  =h_{k}^{0}$. These equations can be integrated to
give
\begin{equation}
h_{k}(\tau)=h_{k}^{0}+(h_{k}^{0}-g_{k}^{0})(k+1) \left\{  \sqrt{\frac
{kT}{kT+\tau}\frac{( k+1) T+\tau}{(k+1) T}}-1\right\}  \label{hk(tau)}%
\end{equation}%
\[
g_{k}\left(  \tau\right)  =g_{k}^{0}+\left(  h_{k}^{0}-g_{k}^{0}\right)
k\left\{  \sqrt{\frac{kT+\tau}{kT}}\sqrt{\frac{\left(  k+1\right)  T}{\left(
k+1\right)  T+\tau}}-1\right\}  .
\]
In particular, at time $\tau=T$%
\begin{equation}
h_{k}\left(  T\right)  =h_{k}^{0}+\left(  h_{k}^{0}-g_{k}^{0}\right)  \left\{
\sqrt{k\left(  k+2\right)  }-k-1\right\}  \label{hk_recurrence}%
\end{equation}%
\begin{equation}
g_{k}\left(  T\right)  =g_{k}^{0}+\left(  h_{k}^{0}-g_{k}^{0}\right)
k\left\{  \frac{k+1}{\sqrt{k\left(  k+2\right)  }}-1\right\}  .
\label{gk_recurrence}%
\end{equation}

It can be verified that $\alpha_{k}\left(  T\right)  =\alpha_{k+1}^{0}%
,h_{k}\left(  T\right)  =h_{k+1}^{0}$ and $g_{k}\left(  T\right)  =g_{k+1}%
^{0}$, thus ensuring the regeneration of the curve $\Gamma$ after a period
$T$. Special attention should be given to the \ boundary region $k=0$. The
equation for $g_{0}\left(  \tau\right)  $ reads%
\[
\frac{-x_{0}\left(  \tau\right)  }{g_{0}\left(  \tau\right)  -\frac{1}{2}%
}=2\left(  T+\tau\right)
\]
Multiplying by denominator and differentiating with respect to $\tau$ we
obtain $2\left(  \tau+T\right)  \frac{\partial g_{0}\left(  \tau\right)
}{\partial\tau}+2\left(  g_{0}\left(  \tau\right)  -\frac{1}{2}\right)
=-\frac{\partial x_{0}\left(  \tau\right)  }{\partial\tau}$. At point
$x_{0}\left(  \tau\right)  $ there is a jump, $\rho\left(  x_{0}\left(
\tau\right)  -0,\tau\right)  =g_{0}\left(  \tau\right)  $ and $\rho
(x_{0}\left(  \tau\right)  +0,\tau)=1$ at the right, consequently

$\frac{\partial x_{0}}{\partial\tau}=\frac{-g_{0}\left(  \tau\right)  \left(
1-g_{0}\left(  \tau\right)  \right)  }{1-g_{0}\left(  \tau\right)  }%
=-g_{0}\left(  \tau\right)  $. Substituting, we obtain
\[
2\left(  \tau+T\right)  \frac{\partial g_{0}\left(  \tau\right)  }%
{\partial\tau}+g_{0}\left(  \tau\right)  -1=0.
\]
Solving the latter with the initial condition $g_{0}\left(  0\right)
=g_{0}^{0}$, we get
\begin{equation}
g_{0}\left(  \tau\right)  =1+\left(  g_{0}^{0}-1\right)  \sqrt{\frac{T}%
{T+\tau}}, \label{g0(tau)}%
\end{equation}
describing the shock propagation during $0\leq\tau\leq T$. Alongside, the jam
joining the boundary will start to dissolve by rarefaction wave $\rho
(x,\tau)=\frac{1}{2}-\frac{x}{2\tau}$ during the green light period $0\leq
\tau\leq\frac{T}{2}$. During the red light period $\frac{T}{2}\leq\tau\leq T$
new jam appears at the boundary $x=0$ and propagates inside. We shall denote
its coordinate by $x_{G}\left(  \tau\right)  $, $x_{G}\left(  \frac{T}%
{2}\right)  =0$. The base of the new jam, denoted by $G\left(  \tau\right)  $,
$G\left(  \frac{T}{2}\right)  =\frac{1}{2}$ will obey
\[
\frac{-x_{G}\left(  \tau\right)  }{G\left(  \tau\right)  -\frac{1}{2}}=2\tau.
\]
Using $\partial x_{G}\left(  \tau\right)  /\partial\tau=-G\left(  \tau\right)
$, we obtain an equation $2\tau\frac{\partial G\left(  \tau\right)  }%
{\partial\tau}+G\left(  \tau\right)  -1=0$, solved by \ $G\left(  \tau\right)
=1-\frac{1}{2}\sqrt{\frac{T/2}{\tau}}$\ \ for $\frac{T}{2}\leq\tau\leq T$. At
time $\tau=T$ one has $G\left(  T\right)  =1-\frac{1}{2\sqrt{2}}=\frac{1}%
{2}+\lambda_{c}=g_{0}^{0}$, restoring the initial shape at $\tau=0$, see
(\ref{g_0}). $g_{1}^{0}$ is determined from the solution (\ref{g0(tau)}),
$g_{0}\left(  T\right)  =g_{1}^{0}$. At time $\tau=T$ the jam, which was
initially at position $x_{0}^{0}$, reaches point $x_{1}^{0}$, consequently
$h_{1}^{0}=1$. The remaining relations (\ref{h_k})-(\ref{x_k}) are obtained
recurrently from (\ref{hk_recurrence},\ref{gk_recurrence}). Thus the proof of
periodicity of the sawtooth structure is completed.

\subsection{Steady state selection}

In the previous subsection we have proved stationarity of the saw tooth state
under periodic driving, but we did not address the question whether this
stationary state is actually reached for any initial state characterized by
the initial density $\lambda$. In order to investigate this problem of steady
state selection we consider an initial state consisting of homogeneous state
on the left $\rho_{-}(x)=\lambda$ and sawtooth structure on the right, joined
by a shock, and demonstrate that it is analogous to a shock between the two
homogeneous states $\rho_{-}(x)=\lambda$ and $\rho_{+}\left(  x\right)
=1-\lambda_{c}$.

The latter shock moves with the velocity $v=\left(  j\left(  \lambda\right)
-j\left(  1-\lambda_{c}\right)  \right)  /\left(  1-\lambda-\lambda
_{c}\right)  =\lambda_{c}-\lambda$, and will travel a distance $\left(
\lambda_{c}-\lambda\right)  T$ after time $T$. In particular, the shock will
be stationary for $\lambda=\lambda_{c}$. Let us prove this feature for the
shock between the homogeneous state $\rho_{-}(x)=\lambda_{c}$, for $x<x_{n}$
and sawtooth structure with $n$ complete sawteeth $\rho_{+}(x)=\Gamma$, for
$x_{n}<x<0$. Heights of all sawteeth $h_{k}$, $k\leq n$ will satisfy
(\ref{h_k_eq_ini}), but the velocity for the $n$-th sawtooth satisfies
\begin{equation}
\frac{\partial x_{k}}{\partial\tau}=\frac{h_{k}(\tau)(1-h_{k}(\tau
))-\lambda_{c}(1-\lambda_{c})}{h_{k}(\tau)-\lambda_{c}}=1-h_{k}(\tau
)-\lambda_{c}. \label{dxn/dt}%
\end{equation}
Multiplying (\ref{h_k_eq_ini}) for $k=n$ by $h_{n}\left(  \tau\right)
-\frac{1}{2}$, differentiating with respect to $\tau$ and using (\ref{dxn/dt})
we obtain $2\left(  \tau+nT\right)  \frac{\partial h_{n}\left(  \tau\right)
}{\partial\tau}+h_{n}\left(  \tau\right)  -\lambda=0$, whose solution is
$h_{n}\left(  \tau\right)  =\lambda_{c}+\left(  h_{n}^{0}-\lambda_{c}\right)
\sqrt{\frac{nT}{nT+\tau}}$. We need to prove that the shock will return to the
original position one after time $T$, i.e. $\int_{0}^{T}\frac{\partial x_{n}%
}{\partial\tau}d\tau=0$. Substituting the solution for $h_{n}\left(
\tau\right)  $ into (\ref{dxn/dt}), and integrating over the period, we have%
\begin{equation}
\int_{0}^{T}\frac{\partial x_{n}}{\partial\tau}d\tau=(2\lambda_{c}%
-1)T+2T(h_{n}^{0}-\lambda_{c})(\sqrt{n(n+1)}-n)=0. \label{hn_stationaryshock}%
\end{equation}
Substitution (\ref{h_k}) satisfies the above equation.

Analogously, one proves that the shock between $\lambda\neq\lambda_{c}$ and
the sawtooth structure, after period $T$, will advance if $\lambda>\lambda
_{c}$ or retreat towards the boundary if $\lambda<\lambda_{c}$. The shock
position after time $T$ is determined by a balance equation. Namely, the extra
mass gained by the shock, $\Delta M=\int_{-\infty}^{0}\left(  \rho\left(
x,T\right)  -\rho\left(  x,0\right)  \right)  dx$, \ is equal to the
difference between the ingoing and outgoing currents $\left(  j_{in}%
-j_{out}\right)  T=\left(  j\left(  \lambda\right)  -j\left(  1-\lambda
_{c}\right)  \right)  T$,%
\[
\Delta M=\left(  j_{in}-j_{out}\right)  T=\left(  j\left(  \lambda\right)
-j\left(  1-\lambda_{c}\right)  \right)  T.
\]
The explicit calculations are not illuminating and are omitted for brevity. As
a guide to the eye, see Fig.\ref{Fig_DentiMF}. The most rapid progression of
the sawtooth structure is achieved for $j_{in}=j_{\max}=1/4$ for which the
number of sawteeth $n$ increases roughly by two each five cycles, see
Fig.\ref{Fig_DentiMF}.

\begin{figure}[ptb]
\begin{center}
\includegraphics[
height=2.2935in, width=2.6645in ]{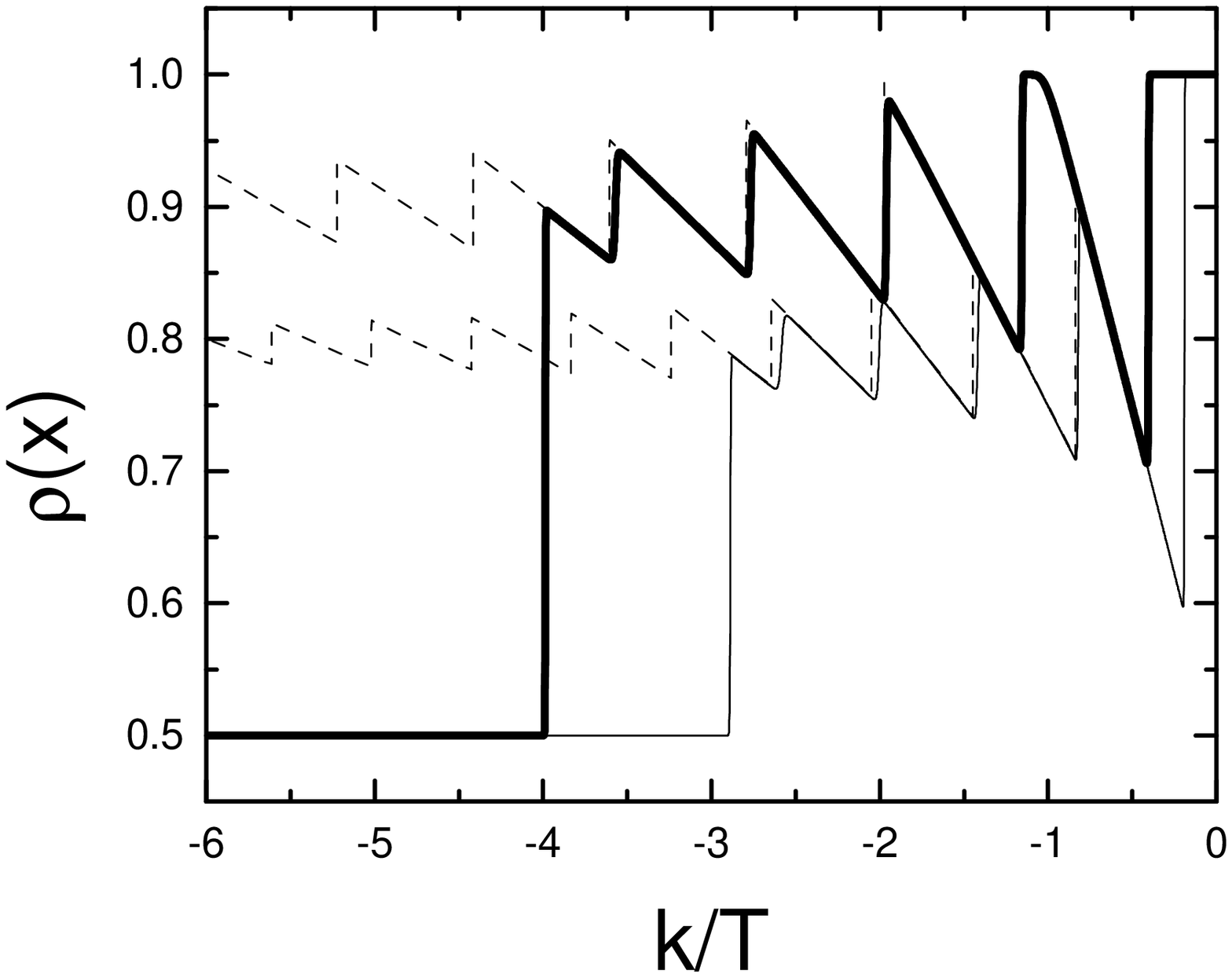}
\end{center}
\caption{Sawteeth density profiles for different fractions of green signal
$f=\tau_{green}/T=0.35,0.65$ Thick ($f=0.35$) and thin ($f=0.65$) curve show
respective density profiles $\rho(x,\tau)$ from mean field equations
(\ref{MeanfieldEq}) evolving from the initial condition $\rho(x,0)=0.5$ after
$10$ full periods. In MF calculations, $\nu=T^{-1}=0.001$. Broken lines show
the limiting curves $\Gamma\left(  f\right)  $, exact in the limit
$\nu\rightarrow0$. The sawteeth heights are given by (\ref{hn(f)}). All the
curves are shown at the moment of time when the red light turns off. }%
\label{Fig_Gam_f}%
\end{figure}

The variations of the boundary density considered in this paper so far
(Eq.(\ref{BC_heaviside})) followed a square wave in which the signal is green
for a fraction $f=\tau_{green}/T=\frac{1}{2}$ of the period. What happens if
we vary the fraction $f$ $\ $between $0$ and $1$? While the cases $f=0$ and
$f=1$ are obvious and were discussed before Eq(\ref{lim_smallnu}), the general
case $0<f<1$ produces a sawteeth structure $\Gamma(f)$, qualitatively similar
to the one for $f=1/2$, compare Figs.\ref{Fig_DentiOrigin},\ref{Fig_DentiMF}
and Fig.\ref{Fig_Gam_f}. The limiting curve $\Gamma(f)$ converges in the bulk
to the value $1-\lambda\left(  f\right)  =\left(  1+\sqrt{1-f}\right)  /2$,
determined by the outgoing flux in the vanishing frequency limit $\langle
j_{out}(f)\rangle=fj_{\max}$, see the discussion after Eq(\ref{jout=jmax/2}).
Note that $\lambda\left(  f\right)  $ satisfies $\lambda\left(  f\right)
\left(  1-\lambda\left(  f\right)  \right)  =\langle j_{out}(f)\rangle=f/4$.
Analytic analysis for arbitrary $f$ can be carried out analogously to the
$f=1/2$ case. In particular, the heights of the sawteeth $h_{n}^{0}\left(
f\right)  $ are determined by the solution of the Eq.(\ref{hn_stationaryshock}%
) where $\lambda_{c}$ is substituted by $\lambda\left(  f\right)  =\left(
1-\sqrt{1-f}\right)  /2$,%
\begin{equation}
h_{n}^{0}\left(  f\right)  =\frac{1}{2}+\frac{\sqrt{1-f}}{2}\sqrt{1+\frac
{1}{n}}\text{,}\label{hn(f)}%
\end{equation}
valid for $n>0$ such that $h_{n}^{0}\left(  f\right)  \leq1$. This
is always the case if $f\geq1/2$ ( duration of green signal is
larger then the duration of the red signal). In the opposite case,
$f<1/2$, additional plateau appear with the saturated density
$\rho=1$ on the curve $\Gamma(f)$, apart from the very first one
caused by the red traffic light at the boundary, see upper curve
in Fig.\ref{Fig_Gam_f}. A thorough analysis of these more
complicated structures is beyond the scope of the present paper.
As a guide for an eye, see Fig.\ref{Fig_Gam_f}, where the curves
$\Gamma\left(  f\right) $ for $f=0.25,0.5,0.65$ are shown.

Let us stress once more the universality aspect of the limiting curve $\Gamma$
shown in Figs.\ref{Fig_Denti_Envelope},\ref{Fig_DentiOrigin}. The shape of the
curve is independent on $T$ provided that $T$ is sufficiently large
$T^{-1}=\nu\rightarrow0$. It also does not depend also on ASEP bulk rates
$p,q$ provided that $p>q$ (drive towards the right boundary), since $\Gamma$
is described by the equation (\ref{InviscidBurgersEq}). Qualitatively,
$\Gamma$ does not depend on relative duration of the green light period with
respect to the whole period $T$ (equal to $1/2$ in the present study), as
argued in the previous paragraph. Moreover, for other models with the convex
current-density relation $j(\rho)$ and traffic light boundary conditions we
expect the existence of a similar curve with sawtooth structure, with
model-dependent shape of the sawteeth. The distance between nearest sawteeth
is determined by $j(\rho)$, see (\ref{delta}). The sawteeth curve in the bulk
will converge to a value $\rho_{R}^{eff}\left(  \nu\right)  $, determined by
averaged outgoing flux through $\langle j_{out}\rangle=j\left(  \rho_{R}%
^{eff}\left(  \nu\right)  \right)  $. The outgoing flux  in the limit
$\nu\rightarrow0$ will be given by (\ref{jout=jmax/2}).

\begin{figure}[ptb]
\begin{center}
\includegraphics[
height=2.4017in, width=3.3155in ]{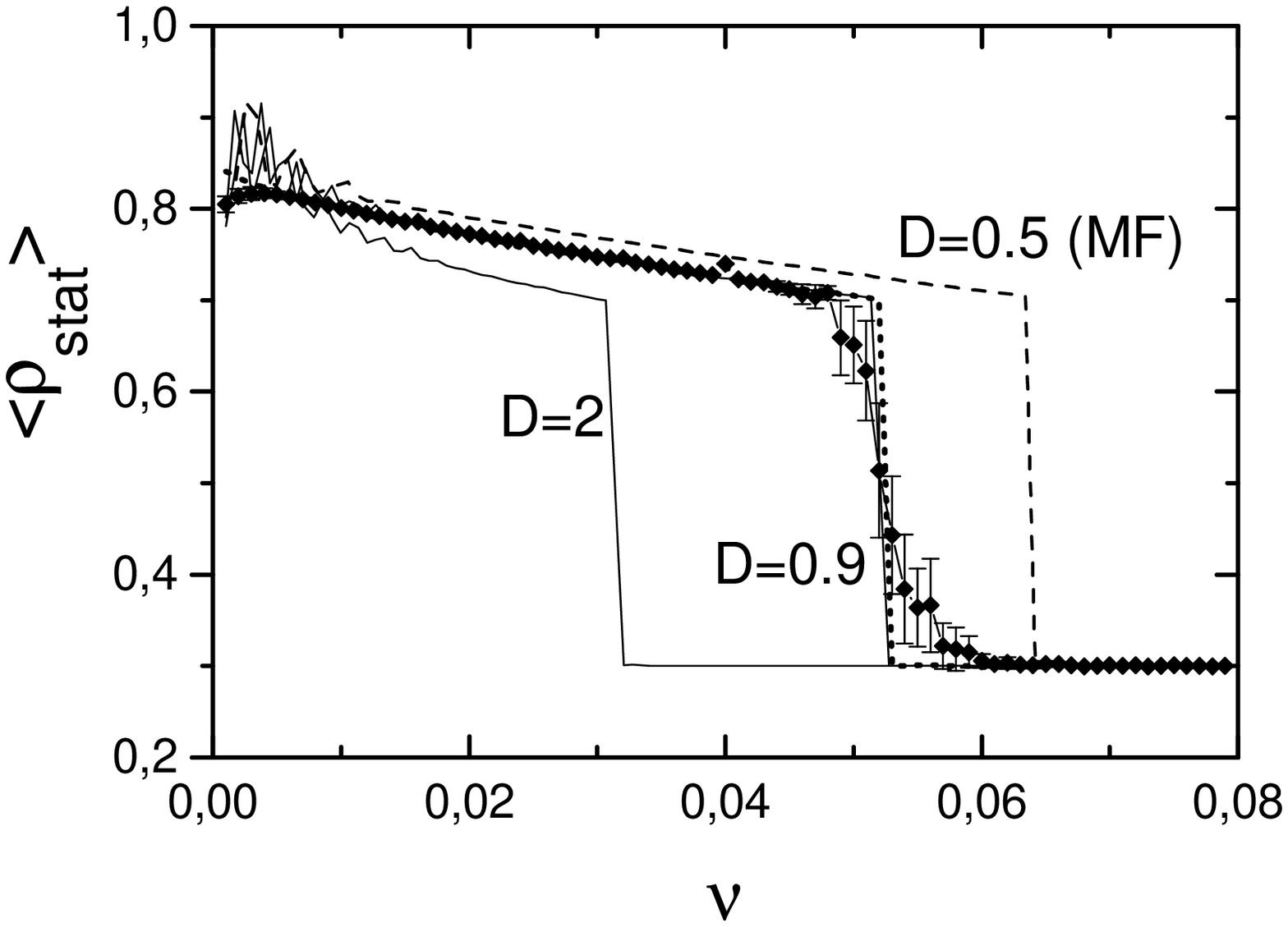}
\end{center}
\caption{Stationary density as function of the frequency $\nu$ for fixed left
boundary density $\rho_{L}=0.3$, from Monte Carlo simulations (diamonds with
errorbars), and from integration of Burgers equation (\ref{BurgersEqD}) with
$D=2,0.9$ (solid lines) and $D=0.5$ (broken line). The dotted line shows
$\langle\rho_{stat}\rangle$ computed from stationary flux measurements.
Fluctuations at small frequencies are due to finite size effects.}%
\label{Fig_nu03_D09}%
\end{figure}

\section{Finite size ASEP\ with traffic light boundary conditions at one or
both boundaries.}

\label{Sec4}

Here we discuss stationary behaviour of sufficiently large but finite system
of size $N$ with open boundaries. It is intuitively clear that the
periodically changing conditions at a boundary will generate a sawteeth
structure with a typical sawtooth size $\Delta\left(  \nu\right)  \approx
\frac{1}{\nu}\left(  \frac{\partial j}{\partial\rho}\right)  _{\rho=\rho
_{R}^{eff}\left(  \nu\right)  }$. We shall consider the case $N\gg\Delta$.
\ If $N\lesssim\Delta$, then one boundary will influence another one during a
periodic cycle.

Firstly, recall the well-known results for the TASEP\ \ model with
time-independent boundary rates, solved exactly in \cite{Schu93,Derr93}. In
the TASEP, a particle can be injected at first site $k=1$ from the left
boundary reservoir with the rate $\alpha$ and be extracted from the last site
$k=N$ with the rate $\beta$. This corresponds to coupling at the left with the
reservoir of particles with density $\rho_{L}=\alpha$ and on the right with
the reservoir of particles with the density $\rho_{R}=1-\beta$. In the range
of parameters $0\leq\alpha,\beta\leq1$ the stationary states are characterized
by average homogeneous particle distribution in the bulk with density
$\rho_{stat}\left(  \rho_{L},\rho_{R}\right)  $. The stationary densities obey
an extremal principle for the stationary flux \cite{Popk99},%
\begin{equation}
j_{stat}=\left\{
\begin{array}
[c]{c}%
\min_{\lbrack\rho_{L},\rho_{R}]}j\left(  \rho\right)  \text{ \ \ if }\rho
_{L}<\rho_{R}\\
\max_{\lbrack\rho_{L},\rho_{R}]}j\left(  \rho\right)  \text{ if }\rho_{L}%
>\rho_{R}%
\end{array}
\right.  , \label{jstat}%
\end{equation}
which, being applied for the case of ASEP $j\left(  \rho\right)  =\rho\left(
1-\rho\right)  $, yields three different phases,
\begin{equation}
\text{Low density (LD) }\rho_{stat}=\rho_{L}\text{, \ for }\rho_{L}%
=\alpha<1/2,\rho_{R}<\rho_{L} \label{LD}%
\end{equation}%
\begin{equation}
\text{High density (LD) }\rho_{stat}=\rho_{R}\text{, \ for }\rho_{R}%
>1/2,\rho_{R}>\rho_{L} \label{HD}%
\end{equation}%
\begin{equation}
\text{Max. current (MC) }\rho_{stat}=1/2\text{, \ for }\rho_{L}>1/2,\rho
_{R}<1/2. \label{MC}%
\end{equation}
\begin{figure}[ptb]
\begin{center}
\includegraphics[
height=2.3313in, width=2.3166in ] {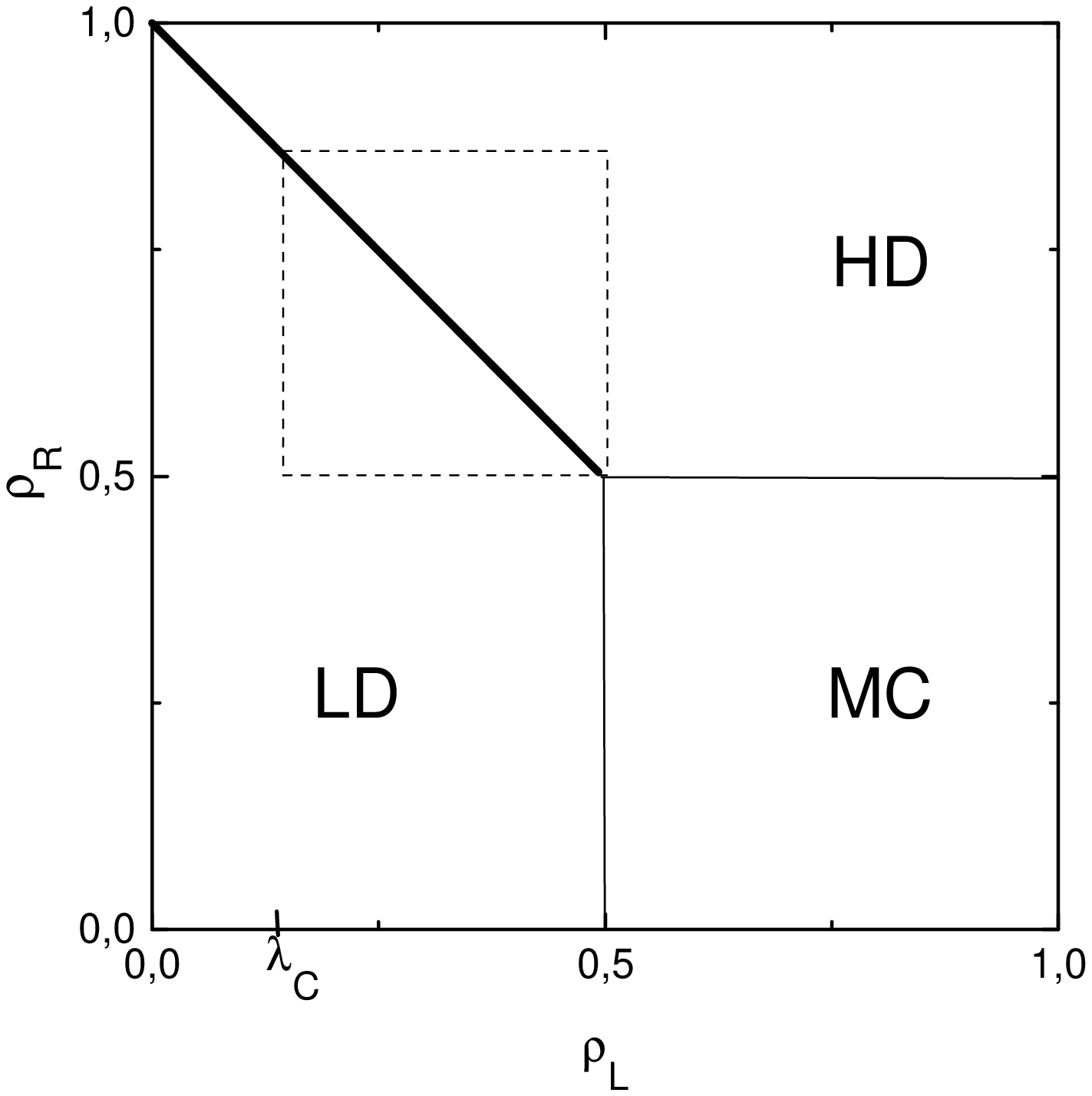}
\end{center}
\caption{ Phase diagram of ASEP with open boundaries, illustrating
(\ref{LD}-\ref{MC}). Solid line indicates discontinuous LD/HD transition.
Dashed line borders the region reachable by applying the traffic light
boundary conditions with arbitrary frequencies at both boundaries. }%
\label{Fig_ASEPdiagram}%
\end{figure}

In the case of fixed left boundary conditions $\rho_{L}=const$ and
periodically changing right boundary conditions (e.g. traffic light boundary
conditions) the $\rho_{R}$ in (\ref{jstat}-\ref{MC}) has to be replaced by the
effective boundary density $\rho_{R}^{eff}\left(  \nu\right)  $ and
$j_{stat},\rho_{stat}$ by a time-averaged flux and density in a stationary
state $\langle j_{stat}\rangle,\langle\rho_{stat}\rangle$ respectively. Note
that since $\rho_{R}^{eff}\left(  \nu\right)  \geq1/2$ for any $\nu$, only LD
and HD phases can be observed.

In particular, (\ref{jstat}-\ref{MC}) with the latter substitution predicts a
discontinuous change of stationary density from $\langle\rho_{stat}%
\rangle=\rho_{L}<1/2$ (LD phase) to $\langle\rho_{stat}\rangle=\rho_{R}%
^{eff}\left(  \nu\right)  $ (HD phase) at a transition point $1-\rho_{L}%
=\rho_{R}^{eff}\left(  \nu\right)  $. Indeed, keeping $\rho_{L}$ fixed and
changing $\nu$, one observes this phase transition at the predicted point, see
Fig.\ref{Fig_nu03_D09}.

Analogously, keeping fixed right boundary conditions $\rho_{R}=const $ and
applying traffic light boundary conditions at the left boundary, one has to
define the effective left boundary density $\rho_{L}^{eff}\left(  \nu\right)
$. The latter, due to particle-hole symmetry of the TASEP, is given simply by
\begin{equation}
\rho_{L}^{eff}\left(  \nu\right)  =1-\rho_{R}^{eff}\left(  \nu\right)  .
\label{particle-hole}%
\end{equation}
Consecutively, the $\rho_{L}^{eff}\left(  \nu\right)  $ varies between
$\rho_{L}^{eff}\left(  \nu\rightarrow+0\right)  =\lambda_{c}$ to $\rho
_{L}^{eff}\left(  \nu\rightarrow\infty\right)  =1/2$.

Finally, applying traffic light boundary conditions at both boundaries with
the frequencies $\nu$ and $\nu^{\prime}$ on the left and on the right,
respectively, effective boundary reservoirs $\rho_{L}^{eff}(\nu) $ at the left
and $\rho_{R}^{eff}\left(  \nu^{\prime}\right)  $ at the right are created.
Again, one finds the phase diagram applying the rule (\ref{jstat}) with the
replacements $\rho_{L}\rightarrow\rho_{L}^{eff}\left(  \nu\right)  $,$\rho
_{R}\rightarrow\rho_{R}^{eff}\left(  \nu^{\prime}\right)  $. Taking into
account (\ref{particle-hole}), one predicts LD (\ref{LD}) phase $\langle
\rho_{stat}\rangle=\rho_{R}^{eff}\left(  \nu\right)  $ for $\nu>\nu^{\prime}$
and HD (\ref{HD}) phase $\langle\rho_{stat}\rangle=\rho_{L}^{eff}\left(
\nu\right)  =1-\rho_{R}^{eff}\left(  \nu\right)  $ for $\nu<\nu^{\prime}$. Due
to the range of variance of the effective boundary densities $\rho_{L}%
^{eff}\left(  \nu\right)  <1/2$, $1/2<\rho_{R}^{eff}\left(  \nu\right)  $, the
maximal current phase (\ref{MC}) cannot be reached except at one point, see
Fig.\ref{Fig_ASEPdiagram}.

In addition to Monte-Carlo simulations, we integrated numerically mean field
equations (\ref{MeanfieldEq}). In the limit of infinitely small frequencies
$\nu\rightarrow0$ the solution of the mean field equations  converges to the
solution of the inviscid Burgers equation, see Figs.\ref{Fig_DentiMF}%
,\ref{Fig_Gam_f}. For finite frequencies $\nu>0$, we observe qualitative
agreement between mean field and Monte-Carlo density profiles, see
Fig.\ref{Fig_stro04}. However, the mean field solution apparently fails to to
predict the exact location of the phase transition frequency, see the curve
with $D=0.5$, marked MF, in Fig.\ref{Fig_nu03_D09}. Indeed, as argued after
Eq.(\ref{BurgersEqD}), meanfield equation can be viewed as a discretization of
viscous Burgers equation with the diffusion coefficient $D=0.5$. The mean
field approach fails quantitatively because it neglects correlations which are
present in the sawtooth state. On the other hand, numerical integration of the
Burgers equation (\ref{BurgersEquation}), keeping the diffusion coefficient
$D$ a free parameter, shows that the the effective boundary density (and
consequently the critical frequency) depends on $D$. Manipulating $D$, one can
obtain a better agreement with the Monte Carlo simulations, see
Fig.\ref{Fig_nu03_D09}. It might seem from the Fig.\ref{Fig_nu03_D09} that the
choice $D=0.9$ fits the Monte-Carlo data well. However, the deviation between
Monte-Carlo data and Burgers equation with $D=0.9$ for small frequencies $\nu$
is substantial and it can be seen by comparing graphs of type
Fig.\ref{Fig_nu03_D09} for different $\rho_{L}$ (not shown).

\section{Conclusions}

\label{Sec5}

We have provided a hydrodynamic description of the semi-infinite ASEP with
traffic light boundary conditions. We find a time-periodic stationary sawtooth
solution which is described in detail. We have also addressed the question of
steady state selection, starting from some initial density $\lambda$. The
picture that emerges is similar to that of the usual ASEP with constant
effective reservoir density that we have determined. Despite the sawtooth
structure of the solution, the time averaged density is at sufficiently
\ large distance from the boundary is given by the $\bar{\rho}(x,t)=\rho
^{eff}$, both in low and high density regime. Indeed, considering the motion
of the shock as an effective \ the single-particle problem in an external
potential \cite{Rako03,Evan03} this observation is reminiscent of the motion
of a Brownian particle in a periodically driven stochastic system
\cite{Dutt03}. Our result shows that effective potentials may arise also in
interacting many-body systems as a result of periodic driving.

Our derivation is based on the ASEP as a specific example, but remains valid
for generic driven diffusive systems with convex current-density relation.
There are several quantitative conclusions that one can draw from the exact
hydrodynamic treatment presented above. Firstly, note that the average
outgoing flux, see (\ref{jout_generic}) $\langle j_{out}\rangle=1/8$ is two
times smaller than the maximal flux $\max j\left(  \rho\right)  =j_{\max
}=1/4.$The relation
\begin{equation}
\langle j_{out}\rangle=\frac{j_{\max}}{2} \label{jout=jmax/2}%
\end{equation}
is not a casual, but a rather generic one: during the red light periods,
$\langle j_{out}\rangle_{red}=0$, and an extended jam at the boundary forms.
During the green light period, $\langle j_{out}\rangle_{green}=j_{\max}$,
because the outflow from a jam is governed by a maximization principle
$\langle j_{out}\rangle_{green}=\max_{\rho\subset\left[  0,1\right]  }j\left(
\rho\right)  =j_{\max}$ \cite{Krug91}. Per full period, one obtains
(\ref{jout=jmax/2}).

Secondly, the distance between neighbouring sawteeth rapidly approaches a
constant, $|x_{k+1}-x_{k}|\approx\left(  1-2\lambda_{c}\right)  T$ for $k\gg
1$, as follows from (\ref{x_k}). The value of the constant has simple physical
origin: maximum and minimums of sawtooth structure $\Gamma$, $h_{k}$ amd
$g_{k}$ approach the effective boundary density value $h_{k},g_{k}%
\approx1-\lambda_{c}\pm O\left(  1/k\right)  =\rho_{R}\pm O\left(  1/k\right)
$. Hence, the velocities of the discontinuities for large $k$ approach the
group velocity $v_{group}\left(  \rho_{R}\right)  =\lim_{k\rightarrow\infty
}\frac{h_{k}\left(  \tau\right)  \left(  1-h_{k}\left(  \tau\right)  \right)
-g_{k}\left(  \tau\right)  \left(  1-g_{k}\left(  \tau\right)  \right)
}{h_{k}\left(  \tau\right)  -g_{k}\left(  \tau\right)  }=\left(  \partial
j/\partial\rho\right)  _{\rho=\rho_{R}}$. By periodicity $\Gamma
(0)=\Gamma\left(  T\right)  $ requirement $x_{k+1}\approx x_{k}-v_{group}%
\left(  \rho_{R}\right)  T$, or
\begin{equation}
\frac{|x_{k+1}-x_{k}|}{T}\approx\left(  \frac{\partial j}{\partial\rho
}\right)  _{\rho=\rho_{R}}=\Delta\text{, } \label{delta}%
\end{equation}
The distance between sawteeth converges monotonically and rapidly to the
predicted value: indeed, as follows from (\ref{x_k}), $\lim_{k\rightarrow
\infty}|x_{k}-x_{0}|/T=(k+\frac{1}{2}-\frac{2\lambda_{c}}{1-2\lambda_{c}%
})\left(  \partial j/\partial\rho\right)  _{\rho=\rho_{R}}$. Consequently, the
sum of all deviations does not exceed $10\%$ of the predicted limiting
distance $\Delta$. Hence, one can measure derivative of the flux $\left(
\partial j/\partial\rho\right)  _{\rho=\rho_{R}}$ directly by measuring the
distance between the sawteeth. Estimate of $\left(  \partial j/\partial
\rho\right)  _{\rho=\rho_{R}}$ from (\ref{delta}) for $k=1$ (the first and the
best-visible sawtooth) induces relative error less than $4\%$.

Finally, the amplitude of density variations $\delta\rho\left(  x\right)  $
can be estimated as $|h_{k}^{0}-g_{k}^{0}|$ at a distance $|x_{k}|\approx
Tk\Delta$. From (\ref{h_k}),(\ref{g_k}) we obtain%
\begin{equation}
\delta\rho\left(  -Tk\Delta\right)  \approx\frac{1-2\lambda_{c}}{2k}=O\left(
\frac{1}{k}\right)  . \label{delta_rho(x)}%
\end{equation}

In most of the paper, a semi-infinite system was considered. The analogy of
the sawtooth structure with a shock and the picture of steady state selection
that emerges allows us to consider also finite systems with two open
boundaries through which particles can enter or leave the system. We argue
that the traffic light boundary condition represents a domain of the full
phase diagram that includes part of the first-order transition between low-
and high-density phase. The maximal-current phase is reached only in the point
where it meets the end of the first order transition line.

\acknowledgments V.P. thanks the IFF, Forschungszentrum Jlich, where a part of
the work was completed, for the hospitality and the University of Salerno for
providing a two years research grant during which this work was performed.
M.S. acknowledges partial support from a MUR-PRIN-2005 initiative
\textit{Transport properties of classical and quantum systems}. G.M.S. thanks
the Department of Physics "E.R. Caianiello" for partial support and for the
kind hospitality received during the initial stages of this work.

%\newpage

%\bibliographystyle{revsymb}
%\bibliographystyle{apsrev}
%\bibliography{ABO}
%Produces the bibliography via BibTeX.

%title = "Derivation of the Leroux system as the hydrodynamic limit
%of a two-component lattice gas, math.PR/0304481 ",

\end{document}